\documentclass{ieeeaccess}
 \pdfoutput=1 
\usepackage{cite}
\usepackage{amsmath,amssymb,amsfonts}
\usepackage{algorithmic}
\usepackage[]{color,graphicx}
\usepackage{textcomp}
\usepackage{bm}
\usepackage{url}
\usepackage{ascmac}

\def\x{{\mathbf x}}

\def\t#1{\tilde{#1}}
\def\bb#1{{\mathbb #1}}
\def\r#1{{\rm #1}}

\usepackage{comment}

\begin{document}
\history{Date of publication xxxx 00, 0000, date of current version xxxx 00, 0000.}
\doi{10.1109/ACCESS.2017.DOI}

\title{Trainable Projected Gradient Detector for Massive Overloaded MIMO Channels: Data-driven Tuning Approach}
\author{
  \uppercase{Satoshi Takabe}\authorrefmark{1}\authorrefmark{2}, \IEEEmembership{Member, IEEE},
  \uppercase{Masayuki Imanishi}\authorrefmark{1},
  \uppercase{Tadashi Wadayama}\authorrefmark{1}, \IEEEmembership{Member, IEEE},
  \uppercase{Ryo Hayakawa}\authorrefmark{3}, \IEEEmembership{Student Member, IEEE},
  and \uppercase{Kazunori Hayashi}\authorrefmark{4}, \IEEEmembership{Member, IEEE}}                
  \address[1]{Nagoya Institute of Technology,
		Gokiso, Showa-ku, Nagoya, Aichi 466-8555, Japan
                }
  \address[2]{%
  		RIKEN Center for Advanced Intelligence Project, Japan
  		Nihonbashi, Chuo-ku, Tokyo 103-0027, Japan
                }
  \address[3]{%
		Graduate School of Informatics, Kyoto University,
                Yoshida-Honmachi, Sakyo-ku, Kyoto 606-8501, Japan
                }
  \address[4]{%
		Graduate School of Engineering, Osaka City University, 
                Sugimoto, Sumiyoshi-ku, Osaka 558-8585, Japan
                }
\tfootnote{A part of this work was presented at IEEE ICC 2019~\cite{TPG}.
This work was partly supported by JSPS Grant-in-Aid for Scientific Research (A) Grant Number 17H01280 (TW), 
Grants-in-Aid for Scientific Research (B) Grant Number 16H02878 (TW) and 19H02138 (TW, KH),
 Grant-in-Aid for Young Scientists (Start-up) Grant Number 17H06758 (ST), and 
 {Grant-in-Aid for JSPS Research Fellow Grant Number 17J07055 (RH)}.
}


\corresp{Corresponding author: Satoshi Takabe (e-mail: s\_takabe@nitech.ac.jp).}

\begin{abstract}
This paper presents a deep learning-aided iterative detection algorithm for massive overloaded
multiple-input multiple-output (MIMO) systems
 where the number of transmit antennas $n$ is larger than that of receive antennas $m$.
Since the proposed algorithm is based on the projected gradient descent method with 
trainable parameters, it is named the trainable projected gradient-detector (TPG-detector).
The trainable internal parameters, such as the step-size parameter, 
can be optimized with standard deep learning techniques, i.e.,  
the back propagation and stochastic gradient descent 
algorithms. This approach is referred to as data-driven tuning, and 
ensures fast convergence during parameter estimation in the proposed scheme.
The TPG-detector mainly consists of matrix-vector product operations 
whose computational cost is proportional to $m n$ for each iteration.
In addition, the number of trainable parameters in the TPG-detector is independent of the number of antennas.
These features of the TPG-detector result in a fast and stable training process and reasonable 
scalability for large systems.
Numerical simulations show that the proposed detector achieves a comparable detection performance
to those of existing algorithms for massive overloaded MIMO channels, e.g., the state-of-the-art 
IW-SOAV detector, with a lower computation cost.
\end{abstract}

{
}

\titlepgskip=-15pt
\maketitle

\section{Introduction}\label{sec_intro}

\PARstart{M}{ultiple-input} 
 multiple-output (MIMO) signal processing
 is an indispensable wireless communication technology 
for achieving increased data transfer rates, enhanced reliability, and improved energy efficiency.
In particular, {\em massive} MIMO systems have been widely studied because they can provide the high spectral efficiency required for upcoming communication technologies such as the 5th generation (5G) wireless network standard~\cite{MUMIMO,Yang}.
Since tens or hundreds of antennas are used in a transmitter and receiver,
signal detection for MIMO channels tends to be a computationally intensive task.
It is thus worth studying practical massive MIMO detection algorithms  
which have both low computational complexity and reasonable bit error rate (BER) performance.

In a down-link massive MIMO channel with mobile terminals,
a transmitter in a base station can have many antennas but
a mobile terminal will have far fewer receive antennas
because of restrictions on the cost, space, and power consumption. 
Such a system is known as an {\em overloaded MIMO} system, in which 
the number of transmit antennas $n$ is larger than that of receive antennas $m$.
Overloaded MIMO communications naturally arise in Internet of Things (IoT) wireless networks, 
i.e., data collection by a base station from a large number of sensor nodes
can also be regarded as an up-link overloaded MIMO system
because the number of 
sensor nodes is typically greater than the number of receive antennas at the base station.

A practical detector for massive overloaded MIMO systems should have not only low computational complexity 
but also reasonable BER performance. However, achieving a reasonable balance between the complexity and
detection performance is challenging.
In fact, most naive MIMO detection algorithms have difficulties with either the detection performance or 
the computational complexity.
For example, the zero-forcing detector and minimum mean square error (MMSE) detector~\cite{Shnidman} 
exhibit poor BER performance for overloaded MIMO channels,
while optimal maximum likelihood (ML) detection is computationally intractable.

Among the detection algorithms for overloaded MIMO systems, 
several approximate ML detectors show a reasonable balance between complexity and performance.
Search-based detection algorithms, such as slab-sphere decoding~\cite{SSD}
and enhanced reactive tabu search (ERTS)~\cite{ERTS}, have been proposed for overloaded MIMO channels.
Although these schemes show excellent detection performance, 
they are still computationally expensive for massive MIMO systems.

Some MIMO detection algorithms can be classified 
into the category of convex optimization-based algorithms.
Fadlallah et al. proposed a MIMO detector based on $\ell_1$-regularized minimization~\cite{Fad1}.
Recently, Hayakawa and Hayashi  proposed an iterative detection algorithm with practical computational complexity 
based on  \emph{iterative weighted sum-of-absolute value} (IW-SOAV) optimization~\cite{IW-SOAV1,IW-SOAV2}.
The IW-SOAV provides a remarkable BER performance in the currently available overloaded MIMO detection algorithms with low computational complexity.

Recently, in the field of sparse signal recovery, 
deep learning techniques have attracted great interest 
because they can significantly improve the signal recovery performances of existing sparse signal recovery algorithms. 
Briefly, by unrolling the signal flow of an 
iterative algorithm, we can obtain a signal-flow graph similar to a feedforward neural network,
where parameters are optimized by back propagation and stochastic gradient 
descent methods.
Gregor and LeCun first proposed such an approach, called the learned iterative shrinkage-thresholding algorithm (LISTA)~\cite{LISTA}, 
which exhibits a sparse signal recovery performance far superior to that of the original ISTA~\cite{ISTA2}.
Ito, Takabe, and Wadayama proposed the \emph{trainable ISTA} (TISTA)~\cite{TISTA, TISTA2} 
which provides significantly faster convergence than ISTA and LISTA. 
Several new algorithms inspired by TISTA have also been developed: 
the TPG decoder for LDPC codes~\cite{LP},
OAMP-net for MIMO systems~\cite{OAMP-net}, DL-OAMP~\cite{CP1,CP2}, and SURE-TISTA~\cite{SURE}.

The emergence of deep learning has also made a great impact on the design of algorithms for wireless communications.
Deep learning-based MIMO detectors, such as the deep MIMO detectors (DMDs) in~\cite{DMD1,DMD2}, have been proposed 
 based on the concept of end-to-end modeling of a detector by a neural network~\cite{Oshea,Wang,Wen}.
Although these algorithms exhibit a reasonable detection performance, 
they have poor scalability because of the large number of tuning parameters, and their computational cost. 
It thus may be difficult to apply them to massive overloaded MIMO systems.

The goal of this paper is to propose a novel detection algorithm which is suitable for massive overloaded MIMO systems. Since the proposed algorithm is 
based on the trainable projected gradient (TPG) algorithm, 
it is called the TPG-detector~\cite{TPG}.
The TPG-detector consists of two iterative steps: 
the gradient descent step and the soft projection step. 
These two steps include several trainable internal parameters
that can be optimized with standard deep learning techniques, i.e.,  
the back propagation and stochastic gradient descent algorithms.
This approach is referred to as data-driven tuning and
ensures the fast convergence of the parameter estimation in the proposed scheme.

This paper is organized as follows. In Section~\ref{sec_ddt}, we introduce the concept of data-driven tuning for iterative algorithms
and demonstrate it with a simple example.
In Section~\ref{sec_set}, we describe the problem setting of massive overloaded MIMO systems.
Section~\ref{sec_tpgd} is the main part of this paper, which introduces the proposed TPG-detector for massive overloaded MIMO systems.
In Section~\ref{sec_nr},
the proposed algorithm's detection performance is compared with other algorithms such as the IW-SOAV.
The last section is devoted to a summary of this paper.
Appendix presents a brief review of the IW-SOAV.

\section{Data-Driven Tuning}\label{sec_ddt}

In this section, we first introduce our key design principle, called data-driven tuning, for numerical optimization algorithms.
A simple example based on a toy problem related to MIMO detection problems 
is then presented to illustrate the basic idea behind data-driven tuning.
In the numerical results, we observe the phenomenon of the data-driven acceleration of convergence for a projected 
gradient descent algorithm. The trainable algorithm shown in the example is then used as the base of the TPG-detector 
proposed in Section \ref{sec_tpgd}.

\subsection{Basic concept}

We here introduce the concept 
of the {\em data-driven tuning of numerical optimization algorithms}, whose origin 
dates back to the work by Gregor and LeCun~\cite{LISTA}.
They unfolded an iterative optimization algorithm and embedded several 
trainable parameters such as matrices in gradient steps
to improve its convergence performance.
In general, by unfolding the iterative processes  (Fig.~\ref{signalflow} (a)) 
and by adding some trainable parameters, we obtain
a  multilayer signal-flow graph that is similar to a deep feedforward neural network (Fig.~\ref{signalflow} (b)).

\Figure[!t]()[width=0.95\columnwidth]{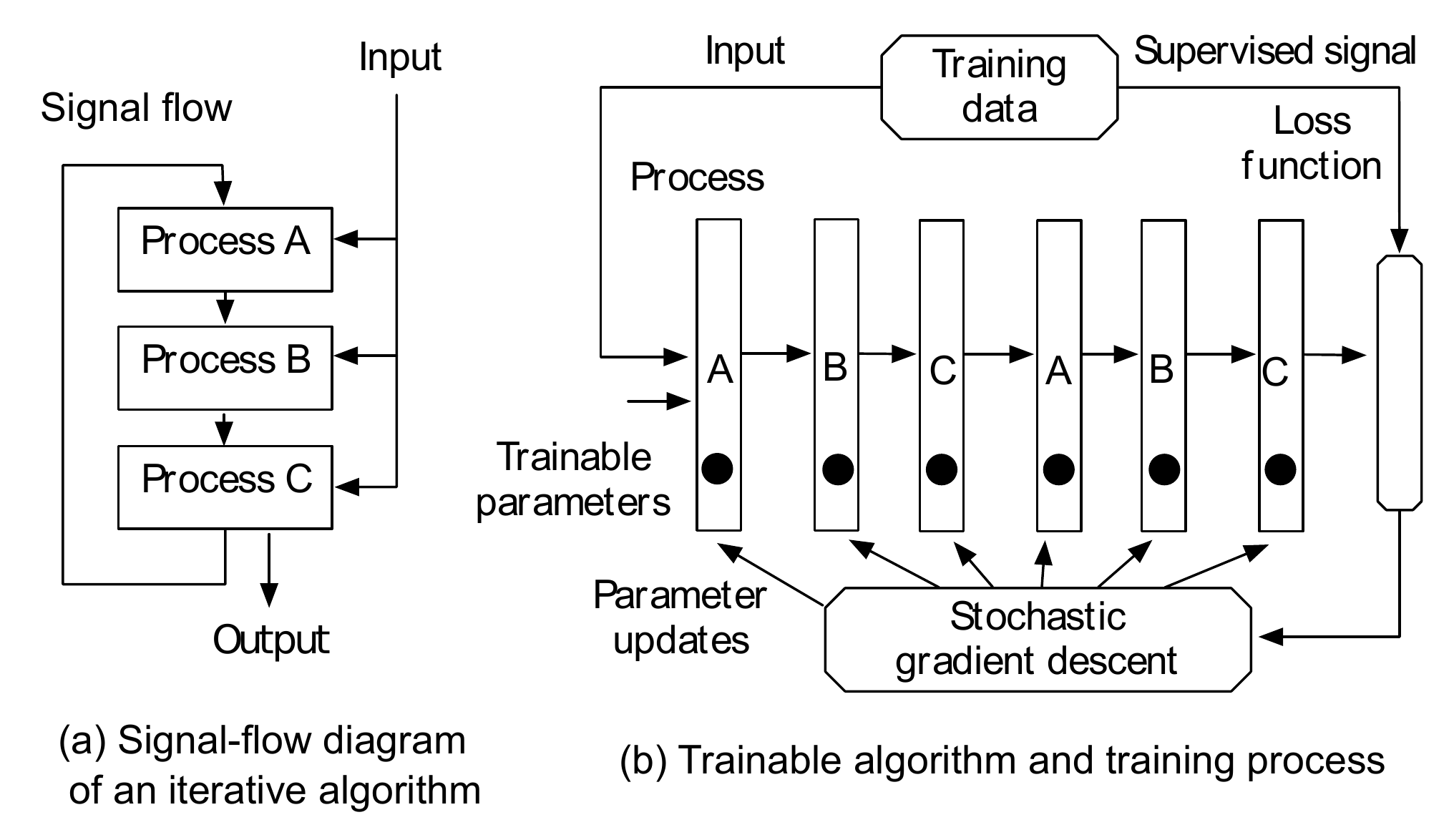}
{(a) A signal-flow diagram of an iterative algorithm, (b) Data-driving tuning based on an unfolded signal-flow graph 
by supervised learning.
 \label{signalflow}}

It is expected that the behavior of each process is controlled by the trainable parameters (black circles in Fig. \ref{signalflow}(b)).
If each process of the signal-flow graph is differentiable,
these trainable parameters can be adjusted by standard deep learning techniques.
Trainable parameters are tuned by minimizing the loss function 
between the supervised signal and the output 
at the end of the unfolded signal-flow graph. 
It is necessary to prepare sufficient training data to tune trainable parameters.
Fortunately, in problems involving wireless communications,
the training data can be randomly generated according to a 
channel model.

In the training process, a randomly generated input is fed into the signal-flow graph 
and the corresponding supervised signal is also fed to the loss function (Fig. \ref{signalflow}(b)).
We apply back propagation and 
an SGD type parameter update (SGD, RMSprop, Adam, etc.)
 to optimize the parameters.

\subsection{Example of data-driven tuning}\label{sec_tpg}
The aim of this subsection is to outline the use of data-driven tuning with a toy problem
similar to the MIMO detection problem.

\subsubsection{Problem setting}
Let us consider a simple quadratic optimization problem 
\begin{equation} 
	\mbox{minimize}_{\bm{x} \in \{-1, +1\}^n}\frac{1}{2}	\| \bm{A} \bm{x} - \bm{y} \|_2^2, \label{eq_QPB}
\end{equation}
where  $\bm{A} \in\mathbb{R}^{n\times n}$ is a real-valued matrix and $\| \cdot \|_2$ represents the Euclidean norm.
We assume that $\bm{y}$ is randomly generated based on a linear observation 
$\bm{y} = \bm{A} \tilde{\bm{x}} + \bm{w} \in \mathbb{R}^n$ where
$\tilde{\bm{x}}$ is a random vector uniformly distributed over $\{-1, +1\}^n$ and
$\bm{w}\in \mathbb{R}^n$ is an i.i.d. Gaussian random vector with zero mean and variance $\sigma^2$.
The optimization problem is regarded as an ML estimation problem for the Gaussian 
linear vector channel, which belongs to NP-hard problems.

In order to approximately solve the problem efficiently, we consider 
a variant of the projected gradient (PG) algorithm.
The PG algorithm considered here is given by 
\begin{eqnarray}
  \label{eq:pg_1}
  \bm{r}_t &=& \bm{s}_t + \gamma \bm{A}^{\mathsf{T}}(\bm{y} - \bm{A} \bm{s}_t),\\
  \label{eq:pg_2}  
  \bm{s}_{t+1} &=& \tanh\left(\xi \bm{r}_t\right),
\end{eqnarray}
where $t=1,\dots, T$ and $\tanh(\cdot)$ is calculated element-wise.
{In this paper, $\bm{H}^\mathsf{T}$ represents the transpose of matrix $\bm{H}$.}
The initial value is set to $\bm{s}_1=\bm{0}$.

There are two processes in the PG algorithm:
In the gradient descent step~(\ref{eq:pg_1}), 
the vector $\bm{r}_t$ is updated along with 
the steepest descent direction of the objective function, i.e.,
$-\nabla \frac 1 2 \|\bm{A} \bm{x} - \bm{y}\|_2^2 = \bm{A}^{\mathsf T}(\bm{y} - \bm{A} \bm{x})$.
The parameter $\gamma$ corresponds to the step-size parameter which controls the convergence behavior
such as the convergence to a fixed point and the convergence speed.
In the projection step~(\ref{eq:pg_2}), we apply a 
soft projection function $\tanh(\cdot)$ to $\bm{r}_t$ in order to obtain the estimate $\bm{s}_{t+1}$
of the $t$th iteration. 
The soft projection ensures that the estimate takes a real value close to $\pm 1$.
Although we can use a hard projection function onto the binary symbols $\{-1, +1\}$ instead of the soft projection,
the PG algorithm fails to converge to a true signal as indicated in Section \ref{sec_tpg_soft}.
In this projection step, we have the parameter $\xi$ which adjusts the softness of the soft projection.
 Note that this type of nonlinear projection has been commonly used 
in several iterative multiuser detection algorithms such as the soft parallel interference canceller~\cite{PIC}.

\subsubsection{Trainable PG algorithm}\label{sec_tpg_prop}

As described in the last subsection, the trainable algorithm can be constructed based on the 
PG algorithm by unfolding its iterative processes.
We have the architecture of the trainable PG (TPG) algorithm given by
\begin{eqnarray}
  \label{eq:tpg_1}
  \bm{r}_t &=& \bm{s}_t + \gamma_t \bm{A}^{\mathsf{T}}(\bm{y} - \bm{A} \bm{s}_t),\\
  \label{eq:tpg_2}  
  \bm{s}_{t+1} &=& \tanh\left(\xi \bm{r}_t\right),
\end{eqnarray}
with initial condition $\bm{s}_1=\bm{0}$.
In the TPG algorithm, we have
trainable parameters $\{\gamma_t \}_{t=1}^{{T}}$ in the gradient descent step\footnote{In the implementation, an alternative trainable parameter 
$\tilde{\gamma}_t^2$ is tuned instead of $\gamma_t$ to satisfy $\gamma_t=\tilde{\gamma}_t^2\ge 0$. This is also true for the TPG-detector in Section~\ref{sec_det}.}. 
Tuning these parameters also adjusts the step-size at each iteration.
In the following discussion, the parameter $\xi$ is treated as a fixed hyperparameter in Section~\ref{sec_tpg_num}
 and it is treated as a trainable parameter in Section~\ref{sec_tpg_soft}.

The trainable parameters are optimized by the standard mini-batch training.
The $i$th training data point $\bm{d}^{i} \triangleq (\bm{x}^i, \bm{y}^i)$ is randomly generated.
The $i$th input signal $\bm{x}^i \in \{-1, +1 \}^n$ is chosen from the uniform distribution and
the corresponding $\bm{y}^i$ 
is generated according to $\bm{y}^i = \bm{A} \bm{x}^i + \bm{w}^i$ for a given $\bm{A}$.
The training dataset of size $D$,  $\mathcal{D}\triangleq \{\bm{d}^{1}, \bm{d}^{2},\dots, \bm{d}^{D}\}$, 
can be regarded as a mini-batch and is fed into the TPG algorithm simultaneously.
In the following experiment, a matrix $\bm{A}$ is randomly generated for each mini-batch
to simulate a realistic situation in which a channel matrix changes frequently.
Here, each element of $\bm{A}$ follows the zero-mean Gaussian distribution with unit variance.

For the $t$th round of the training process, we feed a mini-batch with $D$ training data points 
to the TPG algorithm to minimize the squared loss function
\begin{equation}
 L(\bm{\Theta}_t) \triangleq D^{-1} \sum_{\bm{d}^i \in \mathcal{D}} \| \bm{x}^i -  \hat{\bm{x}}_t(\bm{y}^i) \|^{2}_{2}, \label{eq:sql}
\end{equation}
where $\hat{\bm{x}}_t(\bm{y}) \triangleq  \bm{s}_{t+1}$ is the output of the TPG algorithm after $t$ iterations and $\bm{\Theta}_t\triangleq (\gamma_1,\dots,\gamma_t)$ 
{(or $\bm{\Theta}_t\triangleq (\gamma_1,\dots,\gamma_t, \xi)$)}
 is a vector containing trainable parameters up to the $t$th iteration.
A back propagation process evaluates the gradient $\nabla   L(\bm{\Theta}_t)$, which is 
used for updating trainable parameters
$\bm{\Theta}_t$ as $\bm{\Theta}_t:= \bm{\Theta}_t + \bm{\Delta}$ where $\bm{\Delta}$ is given 
by an SGD type algorithm such as the Adam optimizer~\cite{Adam}. 

It should be remarked that a simple single-shot training using $t = T$
often fails to tune the training parameters accurately because of
 the vanishing gradient phenomenon.
In the TPG algorithm, the phenomenon is caused by
the fact that the derivative of the soft projection function (\ref{eq:tpg_2}) becomes close to zero almost everywhere.
Figure~\ref{fig:tpg_grad} shows the absolute values of the gradient of the trainable parameters $\{\gamma_t\}_{t=1}^T$ in the TPG algorithm.
We find that the gradient vanishes as the iteration index $t$ becomes small.
{The parameter estimators fail to converge in the single-shot training (see Fig.~\ref{fig:tpg_iter})}. 
In order to improve the accuracy of the training process, we use  
the incremental training, which is effective in TISTA~\cite{TISTA}.
In the incremental training, the parameters $\{\gamma_t\}_{t=1}^T$ are sequentially 
trained from $\bm{\Theta}_1$ to $\bm{\Theta}_T$ in an incremental manner.

The details of the incremental training are as follows. First, $\bm{\Theta}_1$ is tuned by minimizing $L(\bm{\Theta}_1)$.
After finishing the training of $\bm{\Theta}_1$, 
the values of the trainable parameters in $\bm{\Theta}_1$ are copied to 
the corresponding parameters in $\bm{\Theta}_2$.
In other words, the results of the training for $\bm{\Theta}_1$ are carried over to $\bm{\Theta}_2$ as the initial values.
For each round of the incremental training, which is called a {\em generation}, $K$ mini-batches are processed.

\Figure[!t]()[width=0.95\columnwidth]{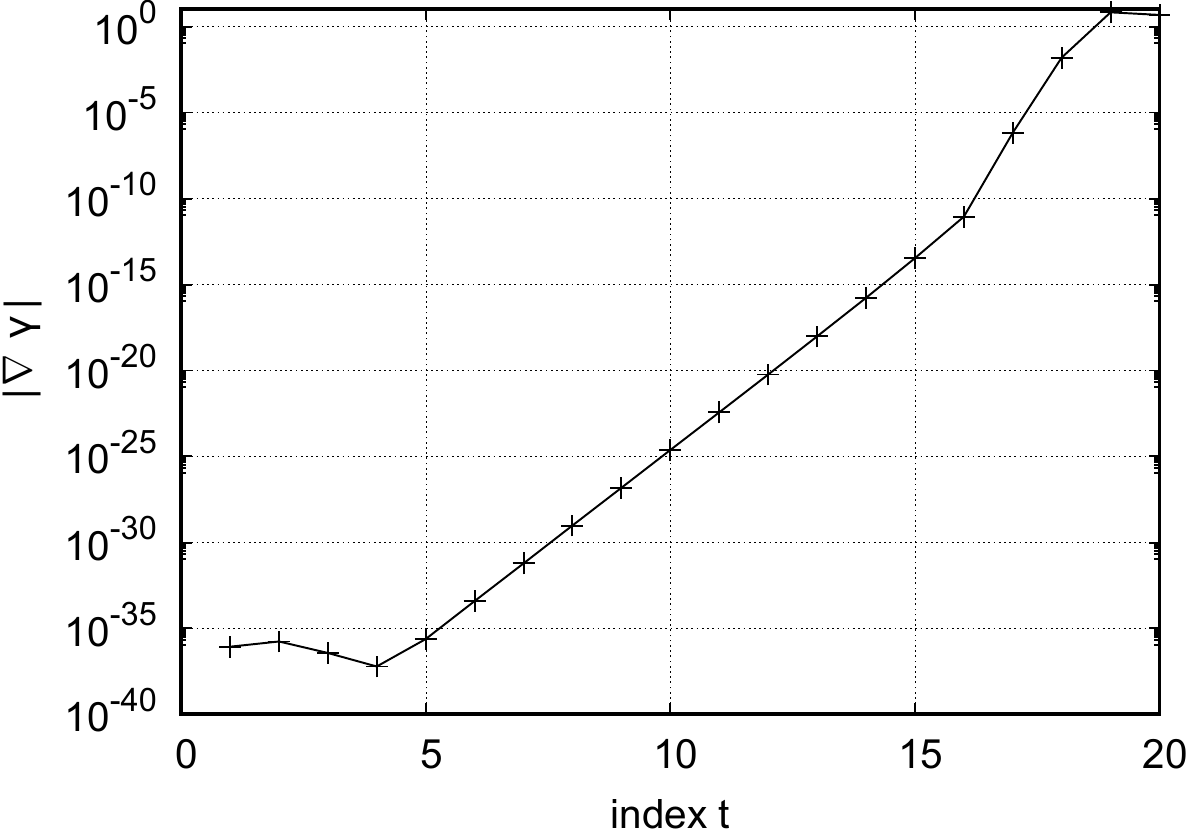}
{Absolute value of gradient of $\{\gamma_t\}_{t=1}^T$
 in the TPG algorithm with single-shot training.
 The results obtained after the 10th mini-batch training by the Adam optimizer with learning rate $2.0\times 10^{-3}$ are
 $n=1000$, $\sigma^2=4.0$, $D=200$, and $\xi=8.0$. \label{fig:tpg_grad}}

\subsubsection{Effect of step-size parameter}\label{sec_tpg_num}

We show the effects of the data-driven tuning in the TPG algorithm.
Here, we treat the step size $\{\gamma_t\}_{t=1}^{{T}}$ as a trainable parameter, while $\xi$ is treated as a hyperparameter.
In the experiment, the dimension of matrix $\bm{A}$ is {set to} $n=1000$ and we set $\sigma^2=4.0$.
The number of iterations of the TPG algorithm is $T = 20$.
We performed two types of training processes for the TPG algorithm to measure the effect of the incremental training.
In the training process with incremental training, we used $K=100$ mini-batches per generation.
The mini-batch size was set to ${D}=200$ and an Adam optimizer learning rate of $2.0\times 10^{-4}$ 
was used.
In the training process without incremental training (named ``TPG-noINC'' in Fig.~\ref{fig:tpg_iter}), 
we used $K=2000$ and ${D}=200$, and the Adam optimizer learning rate was set to $2.0\times 10^{-3}$.
The initial values of the trainable parameters were $\gamma_t=1.0\times 10^{-4}$ ($t=1,\dots,T$).
In this experiment, the softness parameter $\xi$ was set as $8.0$ for the TPG algorithm.

Figure~\ref{fig:tpg_iter} shows the mean squared error (MSE) as a function of the number of iterations 
of the plain PG algorithm when $\xi=6.0$ and $\gamma=6.5\times 10^{-4}$ for
 the TPG algorithms with/without incremental training.
The MSE after $t$ iterations is defined by $10 \log_{10} (\bb{E}[||\bm{x} - {\hat{\bm{x}}_t}(\bm{y})||_2^2]/n)$ (dB)
and estimated from $10^4$ random samples. 
In the plain PG algorithm, we set
 $\gamma=6.5\times 10^{-4}$, which is the optimal value for $T=20$ (see also Fig.~\ref{fig:tpg_gamma}). 

From Fig.~\ref{fig:tpg_iter}, 
it is found that the MSE of the TPG algorithm is remarkably lower than that of 
 the plain PG algorithm.
The MSE of the TPG algorithm is below $-80$ dB at $t = 8$, while the plain PG yields a smaller MSE after $t = 19$.
In particular, the TPG algorithm shows a much faster convergence at $t=9$, indicating that
tuning the trainable parameters leads to fast convergence.
This is an example of the data-driven acceleration of convergence from introducing data-driven tuning. 
Comparing the TPG algorithm with ``TPG-noINC'' in Fig.~\ref{fig:tpg_iter} without incremental training, 
we find that one-shot training fails to tune the trainable parameters accurately
as predicted by Fig.~\ref{fig:tpg_grad}.
This indicates the importance of the incremental training in the data-driven tuning approach.

In Fig.~\ref{fig:tpg_gamma}, {we show the relation between the step-size parameter $\gamma$ 
and the MSE performance of the plain PG algorithm with $\xi = 6.0$.}
{It can be observed that} the parameter $\gamma$ must be selected carefully to obtain appropriate convergence.
In other words, the appropriate $\gamma$ region for fast convergence 
is relatively narrow; specifically, we need to choose an initial estimate within the neighborhood of $6.5 \times 10^{-4}$
to achieve $-100$ dB at $T=200$. 
This means that optimization of the step size is critical even for the plain PG algorithm.
In addition, the TPG algorithm achieves a lower MSE (around $-130$ dB), 
which cannot be achieved by the plain PG algorithm with $200$ iteration steps. 
This fact implies that embedding a step-size 
parameter into each iteration step provides a substantial improvement in 
the speed of convergence and the accuracy of the solution.

\Figure[!t]()[width=0.95\columnwidth]{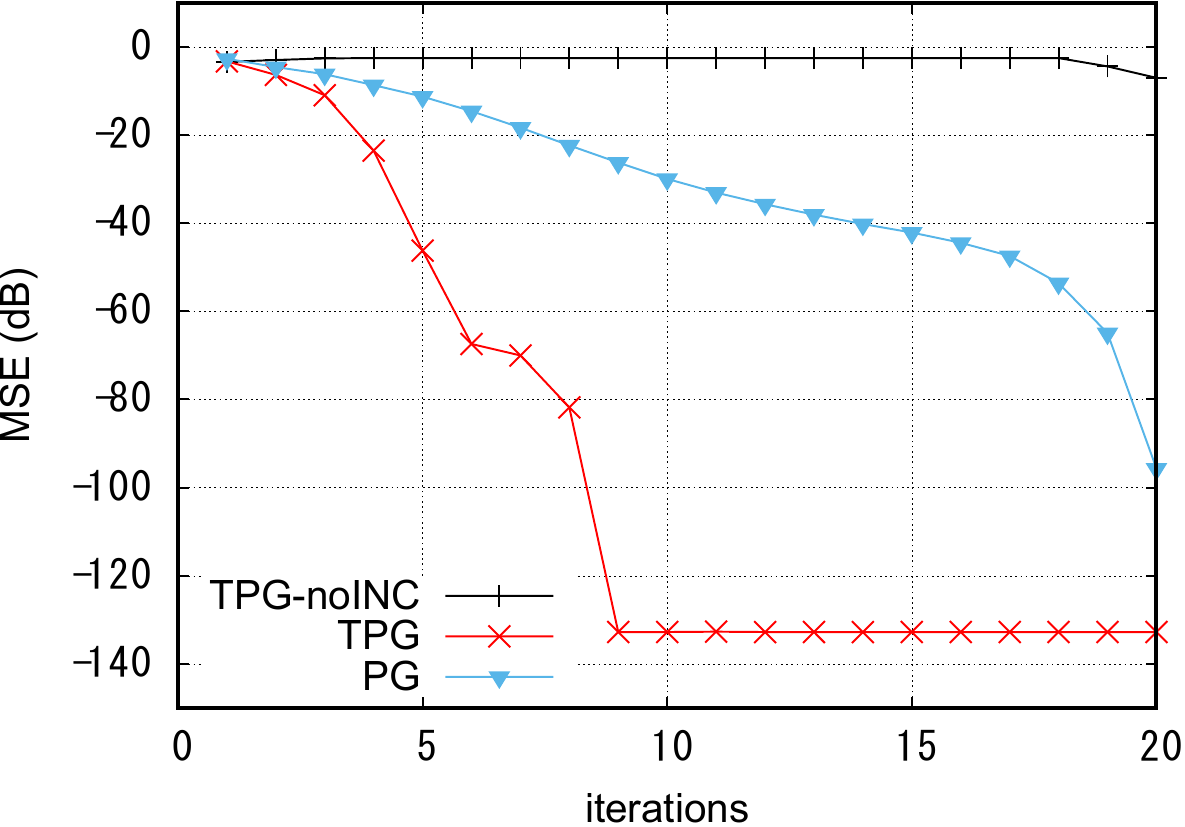}
{MSE as a function of the number of iterations. The curve (PG) represents 
  the MSE of the plain PG algorithm with $\gamma=6.5\times 10^{-4}$ and $\xi=6.0$.
The curve (TPG) corresponds to the MSE of the TPG algorithm and
the curve (TPG-noINC) corresponds to that of the TPG algorithm without incremental training. 
The parameter $\xi$ is fixed at $8.0$ for the TPG algorithm.  \label{fig:tpg_iter}}

\Figure[!t]()[width=0.95\columnwidth]{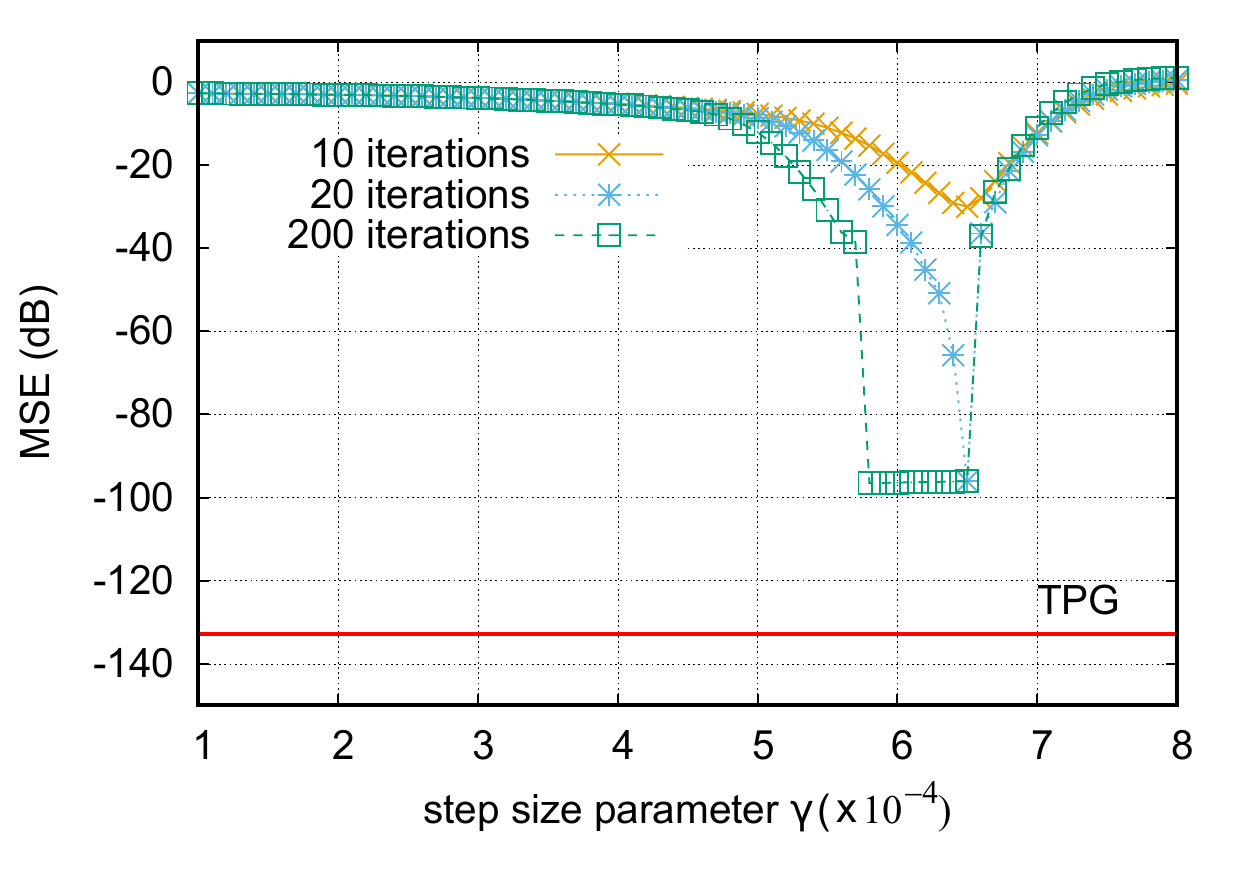}
{Relationship between the step-size parameter $\gamma$ and the MSE performance of the plain PG algorithm with $\xi=6.0$.
  The horizontal solid line represents the MSE of the TPG algorithm with $T=20$ and $\xi=8.0$.
  \label{fig:tpg_gamma}}

\subsubsection{Effect of softness parameter}\label{sec_tpg_soft}
We further discuss the effect of the softness parameter $\xi$ in~(\ref{eq:tpg_2}) {on the MSE performance}.
Figure~\ref{fig:tpg_xitrain} shows the MSE curves of the TPG algorithm with different values of 
 $\xi$ and the TPG algorithm with trainable $\xi$.
The setting of the experiment is the same as above.
As described above, the projection of the TPG algorithm uses the soft projection function
 instead of the hard-projection function corresponding to the $\xi\rightarrow \infty$ limit.
The results show that a large fixed $\xi$ is not appropriate in terms of the MSE.
On the other hand, the TPG algorithm with small fixed $\xi$ also shows a high MSE.
It is thus crucial to tune not only the step-size parameter $\{\gamma_t\}_{t=1}^T$ but also the softness parameter $\xi$ 
{to fully utilize the benefits of the TPG algorithm.
The curve with the label ``TPG ($\xi$ trained)'' represents the MSE of the TPG with trainable $\xi$.
In the experiment, we used $K=10000$ due to the slow convergence of $\xi$.
In Fig.~\ref{fig:tpg_xitrain}, we can see that it outperforms other TPG algorithms with fixed $\xi$.}

\Figure[!t]()[width=0.95\columnwidth]{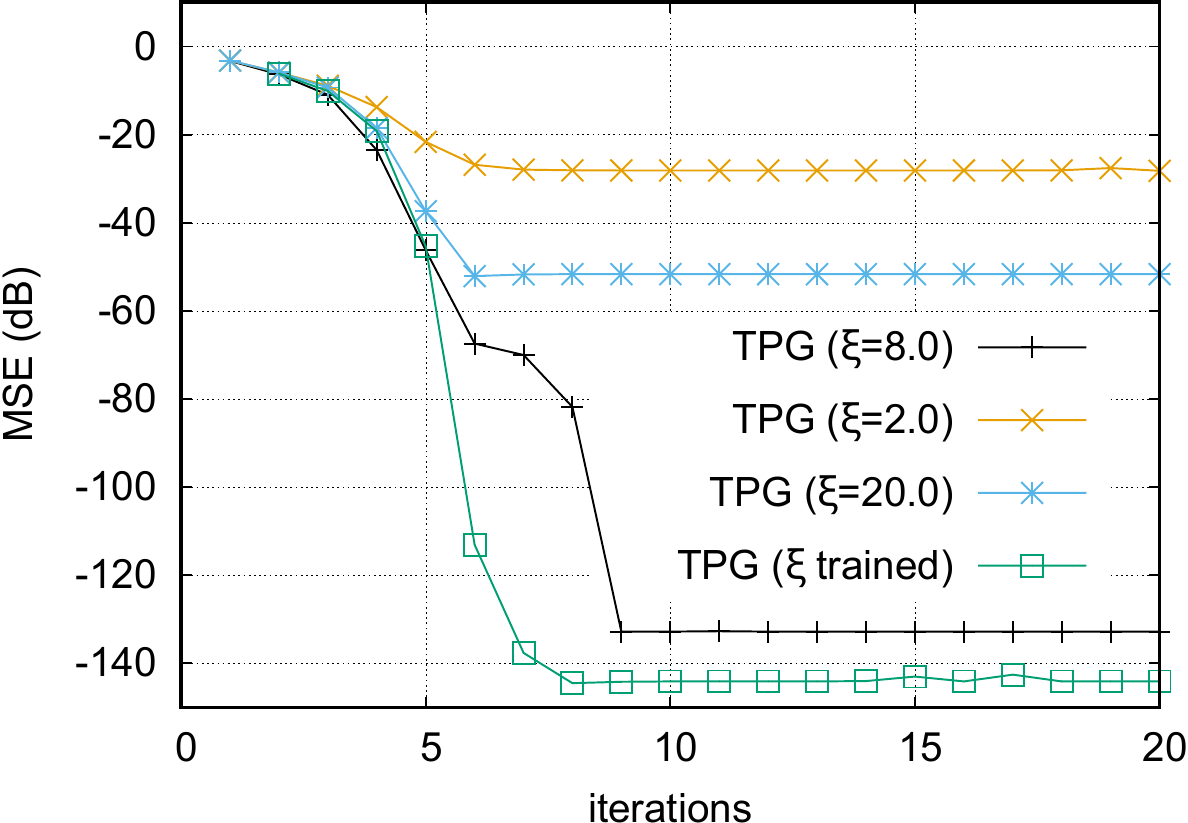}
{MSE as a function of the number of iterations. The curves represent
  the MSE of the TPG algorithm with different fixed $\xi$ or trainable $\xi$.
The trained value is $\xi\simeq 8.560$. \label{fig:tpg_xitrain}}

\subsubsection{Discussion}
From the experimental results shown above, 
it is found that the TPG algorithm shows remarkable acceleration of the convergence speed, which we call 
data-driven acceleration.
{It is emphasized that the optimization problem considered here is randomized because its optimal solution and channel matrix are random variables.
The randomized optimization problem thus has statistical properties in, e.g., gradient information and
landscape of the cost function.}
Data-driven acceleration of convergence is a consequence of data-driven tuning
that aims to learn the stochastic variations on the landscape of the objective functions.
During the training process, trainable parameters are tuned to match the typical objective function.

Most of the known acceleration techniques for gradient descent algorithms,
such as the momentum methods,
do not consider the statistical properties of the problems.
On the other hand, data-driven acceleration does learn the statistical nature of the problem.
The internal parameters controlling the behavior of the algorithm 
are adjusted to match the typical objective function via training processes.
Data-driven acceleration is especially advantageous when implemented in 
detection algorithms because it reduces the number of
iterations required without sacrificing the detection performance. This makes the algorithm
faster and more computationally efficient.

\section{Overloaded MIMO Channels}\label{sec_set}

In this section, we introduce the MIMO channel model used throughout the rest of the paper. 
The numbers of transmit and receive antennas are denoted by $n$ and $m$, respectively.
Our main interest lies in the overloaded MIMO scenario in which $m <n$ holds.
It is also assumed that the transmitter does not use precoding and that the receiver perfectly knows 
the channel state information, i.e., the channel matrix.

Let $\t{\bm{x}} \triangleq [\t{x}_1,\t{x}_2,\dots,\t{x}_n]^{\mathsf{T}} \in \t{{\mathbb S}} ^{n}$ be 
a vector which consists of transmitted symbols $\t{x}_j$ ($j=1,\dots,n$) from the $j$th antenna.
The symbol $\t{\bb{S}} \subset \bb{C}$ represents a signal constellation.
We define 
 $\t{\bm{y}} \triangleq [\t{y}_1,\t{y}_2,\dots,\t{y}_m]^{\mathsf{T}} \in \bb{C}^{m}$ as a vector with
 received symbols $\t{y}_i$ $(i=1,\dots,m)$ by the $i$th antenna.
Assuming a flat Rayleigh fading channel, the received symbol vector $\t{\bm{y}}$ is given by
\begin{equation}
  \t{\bm{y}} = \t{\bm{H}}\t{\bm{x}} + \t{\bm{w}}, \label{eq_1}	
\end{equation}
where $\t{\bm{w}}\in \bb{C}^m$ consists of zero-mean complex Gaussian random variables 
with covariance matrix $\sigma_{w}^2 \bm{I}$.
The matrix $\t{\bm{H}}=(\t{h}_{i,j}) \in \bb{C}^{m \times n}$ is a channel matrix where $\t{h}_{i,j}$ is
the path gain from the $j$th transmit antenna to the $i$th receive antenna.
Each entry of $\t{\bm{H}}$ independently follows a complex circular Gaussian distribution with zero mean and unit variance.
It is convenient to derive an equivalent real-valued channel model 
defined by
\begin{equation}\label{mimo_channnel}
\bm{y} = \bm{H} \bm{x} + \bm{w},	
\end{equation}
where 
    \begin{align} 
      \bm{y} &\triangleq \begin{bmatrix}
        \r{Re}(\t{\bm{y}}) \\
        \r{Im}(\t{\bm{y}})
        \end{bmatrix} \in \bb{R}^{M},\ 
      \bm{H} \triangleq \begin{bmatrix} \label{eq:H_real}
        \r{Re}(\t{\bm{H}}) & - \r{Im}(\t{\bm{H}})\\
        \r{Im}(\t{\bm{H}}) &  \r{Re}(\t{\bm{H}})\\
      \end{bmatrix}, \nonumber\\ 
      \bm{x} &\triangleq \begin{bmatrix}
        \r{Re}(\t{{\bm{x}}}) \\
        \r{Im}(\t{{\bm{x}}})
        \end{bmatrix} \in \bb{S}^{N}, \ 
      \bm{w} \triangleq \begin{bmatrix}
        \r{ Re}(\t{\bm{w}}) \\
        \r{ Im}(\t{\bm{w}})
        \end{bmatrix}\in \bb{R}^{M},
    \end{align} 
 and $(N,M) \triangleq (2n,2m)$. {For $z\in \mathbb{C}$, $\r{Re}(z)$ and $\r{Im}(z)$ 
 represent the real and imaginary part of $z$, respectively.}
The signal set $\bb{S}$ is the real counterpart of $\t{\bb{S}}$.
The matrix $\bm{H} \in \bb{R}^{M \times N}$ is converted from $\t{\bm{H}}$.
 Similarly, the noise vector $\bm{w}$ consists of i.i.d. random variables following a Gaussian distribution
 with zero mean and variance $\sigma^2_w/2$.
The signal-to-noise ratio (SNR) per receive antenna is then represented by
\begin{equation}
  \mathrm{SNR}\triangleq \frac{E_s}{N_0}
  =  \frac{2n}{\sigma_{w}^{2}},
\end{equation}
  where $E_s\triangleq \bb{E}[||\t{\bm{H}}\t{\bm{x}}||^{2}_{2}]/m$ is the signal power per receive antenna and
  $N_0 \triangleq \sigma_w^2$ is the noise power per receive antenna.
Throughout this paper, we assume the QPSK modulation format, i.e., 
$\t{\bb{S}} \triangleq \{1+j,-1+j,-1-j,1-j\}$, 
which is equivalent to the BPSK modulation $\bb{S} \triangleq \{-1,+1\}$
in the corresponding real-valued channel model (\ref{mimo_channnel}).

\section{Trainable Projected Gradient (TPG)-Detector}\label{sec_tpgd}

The proposed algorithm, called the TPG-detector, is based on the TPG algorithm introduced in Section \ref{sec_tpg}.
We first describe the details of the TPG-algorithm and discuss its time complexity.
The key difference between the previously proposed trainable detector~\cite{TPG} 
and the TPG-detector described below is the improvement in 
the gradient step which leads to a significant performance improvement when the ratio $m/n (<1)$ is sufficiently large.
The TPG-detector has a lower computational cost than the OAMP-net~\cite{OAMP-net}
 and requires a smaller number of trainable parameters
than the DMD~\cite{DMD1,DMD2}. These features lead to the low training and execution costs of the TPG-detector.

\subsection{Details of TPG-detector}\label{sec_det}

The ML estimation rule for the MIMO channel is given by
\begin{equation} \label{mlprob}
	\hat{\bm{x}} \triangleq \mbox{argmin}_{\bm{x} \in \{-1, +1\}^N } \| \bm{H} \bm{x} - \bm{y} \|_2^2.
\end{equation}
An exhaustive search for the optimal solution is intractable when the system size is large 
because (\ref{mlprob}) is a non-convex optimization problem.
Similar to Section~\ref{sec_tpg}, we instead approximate the solution by using the TPG algorithm.
The recursive formula of the TPG-detector is given by
\begin{eqnarray}
  \label{eq:hoge_1}
  \bm{r}_t &=& \bm{s}_t + \gamma_t \bm{W} (\bm{y} - \bm{H} \bm{s}_t),\\
  \label{eq:hoge_2}  
  \bm{s}_{t+1} &=& \tanh\left(\frac{\bm{r}_t}{|\theta_t|}\right),
\end{eqnarray}
where $t ( = 1, \ldots, T)$ represents the index of an iteration step (or layer) and we use $\bm{s}_1=\bm{0}$ as the initial value.
The estimate of the algorithm after $T$ iterations is given by $\hat{\bm{x}} = \bm{s}_{T+1}$.

The processes  (\ref{eq:hoge_1}) and (\ref{eq:hoge_2}) correspond to the gradient descent step and
the soft projection step, respectively, with soft projection function $\tanh(\cdot)$.
The matrix $\bm{W}$ in the gradient step (\ref{eq:hoge_1}) is the linear MMSE (LMMSE)-like
matrix defined by
\begin{equation} \label{LMMSE_matrix}
	\bm{W} \triangleq \bm{H}^{\mathsf T} (\bm{H} \bm{H}^{\mathsf T} + \alpha \bm{I} )^{-1},
\end{equation}
where $\alpha \in \mathbb{R}$ is a trainable parameter.
The matrix (\ref{LMMSE_matrix}) also appears in the solution of 
the linear regression problem with a quadratic regularization term.
We note that the OAMP-net~\cite{OAMP-net} uses LMMSE-like matrices as well. 
The key difference from the TPG-detector 
is that the LMMSE-like matrix needs to be computed at each iteration in the OAMP-net.
In fact, in the $t$th iteration step of the OAMP-net, 
the matrix is given as $\bm{W}_t$ by substituting the error variance with $\alpha$ in (\ref{LMMSE_matrix}). 

In our previous work~\cite{TPG}, 
the Moore-Penrose pseudo-inverse matrix $\bm H^{\mathsf{T}}(\bm H\bm H^{\mathsf{T}})^{-1}$ was applied to the matrix $\bm{W}$ 
as in the orthogonal approximate message passing (OAMP) algorithm~\cite{OAMP}
and TISTA~\cite{TISTA, TISTA2}, 
although a naive gradient descent method for (\ref{mlprob}) sets $\bm{H}^{\mathsf T}$ to $\bm W$ as in~(\ref{eq:tpg_1}).
However, as we will see in Fig.~\ref{fig:matrix1}, the TPG-detector 
with the Moore-Penrose pseudo-inverse matrix has a poor detection performance 
when the ratio $m/n$ is relatively large.
Here, we instead use the LMMSE-like matrix (\ref{LMMSE_matrix}) for  $\bm{W}$.

Since it is critical to tune $\alpha$ to obtain a reasonable detection performance, 
the parameter $\alpha$ is also optimized in the training process. 
To reduce the number of trainable parameters and the computational cost, we assume that 
the same value of $\alpha$ is used at each iteration.

\subsection{Trainable parameters}

The trainable parameters of the TPG-detector are {$2T+1$
real scalar variables {$\alpha$, $\{\gamma_t\}_{t=1}^T$, and $\{\theta_t\}_{t=1}^T$}.
The parameters $\{\gamma_t\}_{t=1}^T$ in the gradient step
control the step size of the update.
It should be remarked that similar trainable parameters 
are also introduced in the structure of TISTA~\cite{TISTA, TISTA2}.
The parameters $\{\theta_t\}_{t=1}^T$ {control} the softness of the soft projection 
in (\ref{eq:hoge_2}).
Different from the TPG algorithm, the trainable parameters depend on the iteration index $t$
to increase the degree of freedom of the trainable parameters in the soft projection functions.
The parameter $\alpha$ adjusts the degree of compensation for an ill-conditioned matrix $\bm{H}$.
Apart from $\{\gamma_t\}_{t=1}^T$ and  $\{\theta_t\}_{t=1}^T$, the TPG-detector uses the same parameter $\alpha$
for {all} iterations.

One of the advantages of the TPG-detector is that the number of trainable parameters is small, i.e., $O(T)$, 
and this leads to a fast and stable training process.
The number of trainable parameters in the TPG-detector does not depend on {the number of 
antennas $n$ and $m$}
although the DMD~\cite{DMD1} contains $O(n^2T)$ parameters in $T$ layers.
Since the computational cost for the training process is roughly proportional to the number of 
trainable parameters, the TPG-detector results in a remarkably fast training process, similarly to TISTA~\cite{TISTA2}.

\subsection{Time complexity}

The computational complexity of the TPG-detector per iteration 
is $O(mn)$ because calculating the vector-matrix products $\bm{H} \bm{s}_{t}$ 
and $\bm{W} (\bm{y} - \bm{H} \bm{s}_t)$ takes $O(mn)$ computational steps.
We need to calculate the LMMSE-like matrix $\bm{W}$ with $O(m^3)$ computational steps because the calculation 
involves an matrix inversion. However, the calculation of $\bm{W}$ is required only when $\bm{H}$ changes; i.e., the matrix inversion is not needed for each iteration of a TPG-detector if $\bm{H}$ is constant during the process.
The total computational cost for the TPG-detector (for $T$ iterations) without the initialization process for $\bm{W}$
is thus $O(mnT)$.

Another TISTA-based MIMO detection algorithm named OAMP-net~\cite{OAMP-net}
also uses the LMMSE matrix as a linear estimator.
However, its computational cost is higher than that of the TPG-detector 
because the OAMP-net needs to compute a matrix inversion at each iteration~\cite{OAMP-net}.
The total computational complexity of the OAMP-net with $T$ iteration steps is $O(m^3T)$,
which is larger than that of the TPG-detector.
It should be emphasized that the computational cost affects not only the detection processes
but also to the training processes.

\Figure[!t]()[width=0.95\columnwidth]{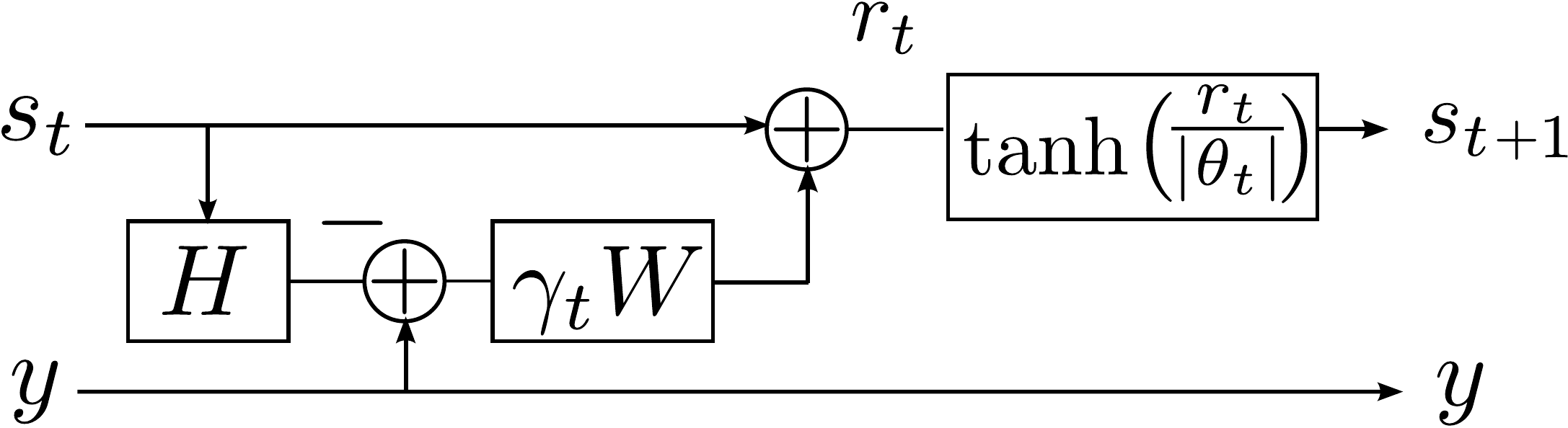}
{The $t$th layer of the TPG-detector. The trainable parameters are $\gamma_t$ and $\theta_t$. \label{fig:TPG-detector}}

\subsection{Training process} 

The TPG-detector is trained based on the incremental training described in Section~\ref{sec_tpg_prop}.
The training data are generated randomly according to the channel model (\ref{mimo_channnel}) 
with fixed variance $\sigma_w^2$ corresponding to a given SNR.
As described in Section~\ref{sec_set}, we assume a practical situation in which a channel matrix $\bm{H}$ is a random variable.
According to this assumption, a matrix $\bm{H}$ is randomly generated for each mini-batch in the training process 
of the TPG-detector.

\section{Numerical Results}\label{sec_nr}

In this section, we present the detection performance of the TPG-detector and compare it to that of other
algorithms such as the IW-SOAV, which is known as one of the most efficient iterative algorithms 
for massive overloaded MIMO systems.

\subsection{Experimental setup}

In numerical experiments, we uniformly sample a
transmitted vector $\bm{x}$
and generate random channel matrices for BER estimation.
The BER is evaluated for a given SNR.

The TPG-detector was implemented with PyTorch 0.4.0 \cite{PyTorch}.
In this paper, a training process is executed with $T=50$ rounds using the Adam optimizer~\cite{Adam}. 
To calculate the BER of the TPG-detector, a sign function $\mathrm{sgn}(z)$ 
which takes a value of $-1$ if $z\le 0$ and $1$ otherwise is applied to the output $\bm{s}_{T+1}$
 in an element-wise manner.

For comparison, we use 
the ERTS~\cite{ERTS}, IW-SOAV~\cite{IW-SOAV2}, and the standard MMSE detector.
The ERTS is a heuristic algorithm based on a tabu search for overloaded MIMO systems.
The parameters of ERTS are based on those given in the original work~\cite{ERTS}.
The computational complexity for executing the main loop of the ERTS algorithm is $O(N_{\mathrm{RTS}}n^2)$
where $N_{\mathrm{RTS}}$ is the maximum number of RTSs which is set to $500$ in this paper.

The IW-SOAV is a double loop algorithm for massive overloaded MIMO systems (see the Appendix for a brief review).
Its inner loop is the W-SOAV optimization, which recovers a signal using a proximal operator.
Each round of the W-SOAV takes $O(mn)$ computational steps, which is comparable to that of the TPG-detector.
After finishing an execution of the inner loop with $K_{\mathrm{itr}}$ iterations,
several parameters are then updated in a re-weighting process based on a tentative recovered signal.
This procedure is repeated $L$ times in the outer loop.
The total number of steps of the IW-SOAV is thus $K_{\mathrm{itr}} L$
and the total computational cost is $O(K_{\mathrm{itr}} Lmn)$ without pre-computation.
In the following, we use the simulation results in~\cite{IW-SOAV2} with $K_{\mathrm{itr}}=50$.
In this case, the computational cost of the TPG-detector is roughly equal to that of the IW-SOAV with $L=1$
if $m/n$ is a constant.
If $L\ge 2$, i.e., the outer re-weighting process is executed, 
and the TPG-detector has lower computational complexity than the IW-SOAV. 

\subsection{Main Results}

\subsubsection{Selection of matrix {$\bm{W}$} in the gradient step}\label{sec_matrix}
As described above, we choose the matrix $\bm{W}$ in the gradient step~(\ref{eq:hoge_1}).
The selection of the matrix will affect the detection performance of the algorithms as shown in the OAMP~\cite{OAMP}.
Before we compare the BER performance of the proposed TPG-detector with other detection algorithms, 
we numerically test {the effect} of the choice of $\bm{W}$.
Figure~\ref{fig:matrix} shows the detection performance of the TPG-detector with a different choice of $\bm{W}$ 
when $(n,m)\!=\!(150,96)$.
{We examined three types of matrices: the matched filter matrix (MF) $\bm{H}^{\mathsf T}$, 
pseudo-inverse matrix (PINV) $\bm H^{\mathsf{T}}(\bm H \bm H^{\mathsf{T}})^{-1}$, and LMMSE-like matrix (LMMSE)~(\ref{LMMSE_matrix}).}
From Fig.~\ref{fig:matrix}, 
we find that the LMMSE-like matrix outperforms other choices of $\bm{W}$ in a wide range of SNR.

\Figure[!t]()[width=0.95\columnwidth]{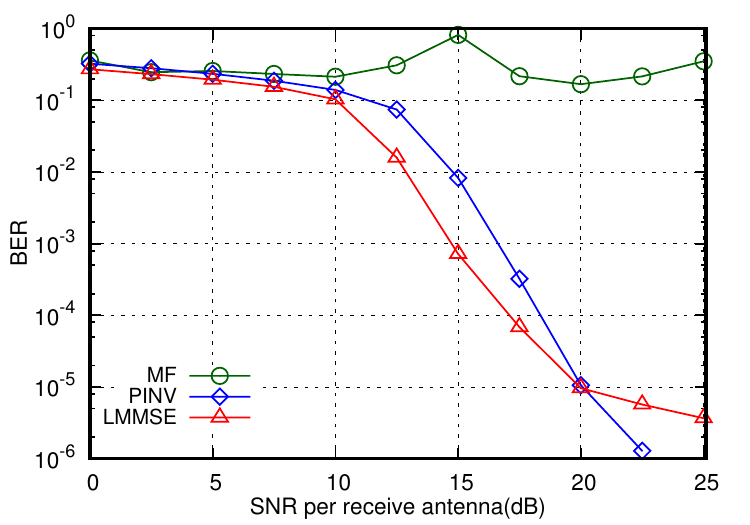}
{BER performances for $(n,m)=(150,96)$ for different {choices} of $\bm{W}$; SNR$=20$ dB.
  {The label ``MF'' stands for the matched filter which corresponds to the gradient of the original problem~(\ref{mlprob}), i.e., $\bm{W}\triangleq \bm{H}^{\mathsf{T}}$, 
``PINV'' represents the pseudo-inverse matrix $\bm{W}\triangleq \bm{H}^{\mathsf T}(\bm{H}\bm{H}^{\mathsf T})^{-1}$, and 
``LMMSE'' is the LMMSE-like matrix~(\ref{LMMSE_matrix}).}
  \label{fig:matrix}  }

\Figure[!t]()[width=0.95\columnwidth]{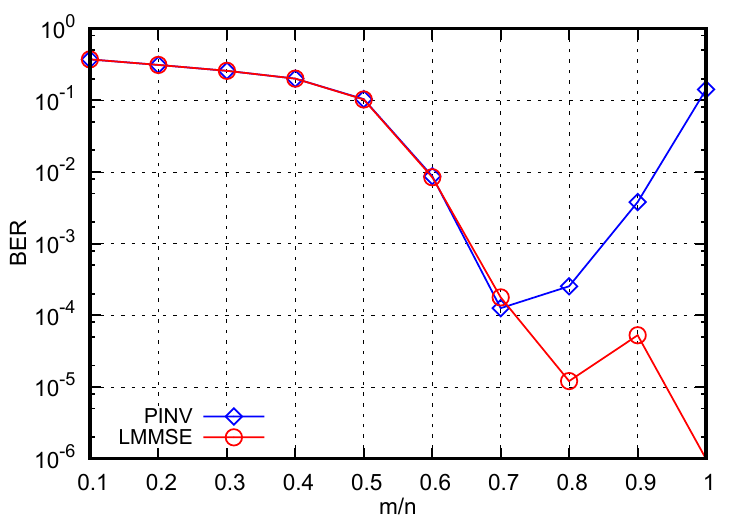}
  {BER performances of the TPG-detector with different {choices} of $\bm{W}$ as a  function of $m/n$; $n=50$, SNR$=20$ dB, $T=50$.
  {The label ``PINV'' represents the pseudo-inverse matrix $\bm{W}\triangleq \bm{H}^{\mathsf T}(\bm{H}\bm{H}^{\mathsf T})^{-1}$ and 
``LMMSE'' is the LMMSE-like matrix~(\ref{LMMSE_matrix}).  }
  \label{fig:matrix1}  }

Figure~\ref{fig:matrix1} shows the BER performances of PINV and LMMSE as a function of the ratio $m/n$ when $n=50$ and SNR$=20$ dB.
Although the BER performance of LMMSE is close to that of PINV
when $m/n\le0.7$, LMMSE shows lower BER values when the ratio $m/n$ is close to one.
This is because the random matrix $\bm{H}$ tends to be ill-conditioned, i.e., 
the condition number of $\bm{H}$ increases as $m/n(<1)$ increases. 
The LMMSE-like matrix successfully compensates the effect of the condition number.
In fact, as shown in Fig.~\ref{fig:matrix2}, the learned value of $\alpha$ becomes large in the high $m/n$ region.
Otherwise, the learned $\alpha$ is close to zero, which suggests that LMMSE  corresponds to PINV without $\alpha$.
From these observations, we set the LMMSE-like matrix to $\bm{W}$ in the proposed TPG-detector
 to treat the ill-conditioned matrix $\bm{H}$.

\Figure[!t]()[width=0.95\columnwidth]{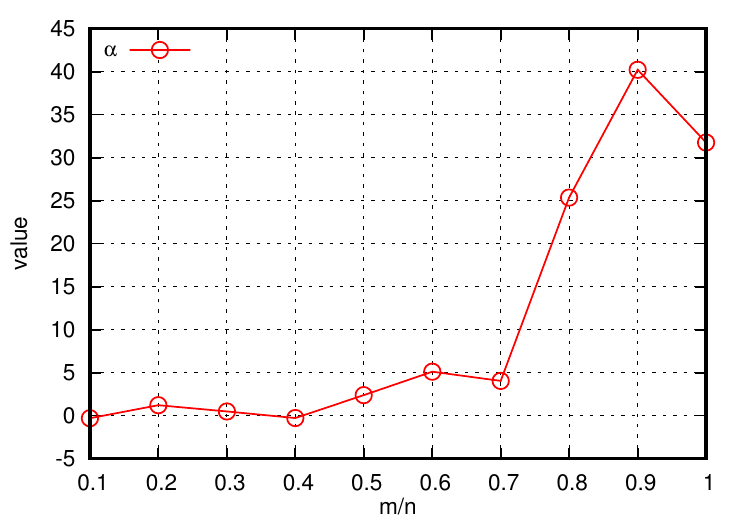}
{The learned values of $\alpha$ in the TPG-detector with the LMMSE-like matrix~(\ref{LMMSE_matrix});
   $n=50$, SNR$=20$ dB, $T=50$. 
  \label{fig:matrix2}  }

\subsubsection{Signal detection process}

\Figure[!t]()[width=0.95\columnwidth]{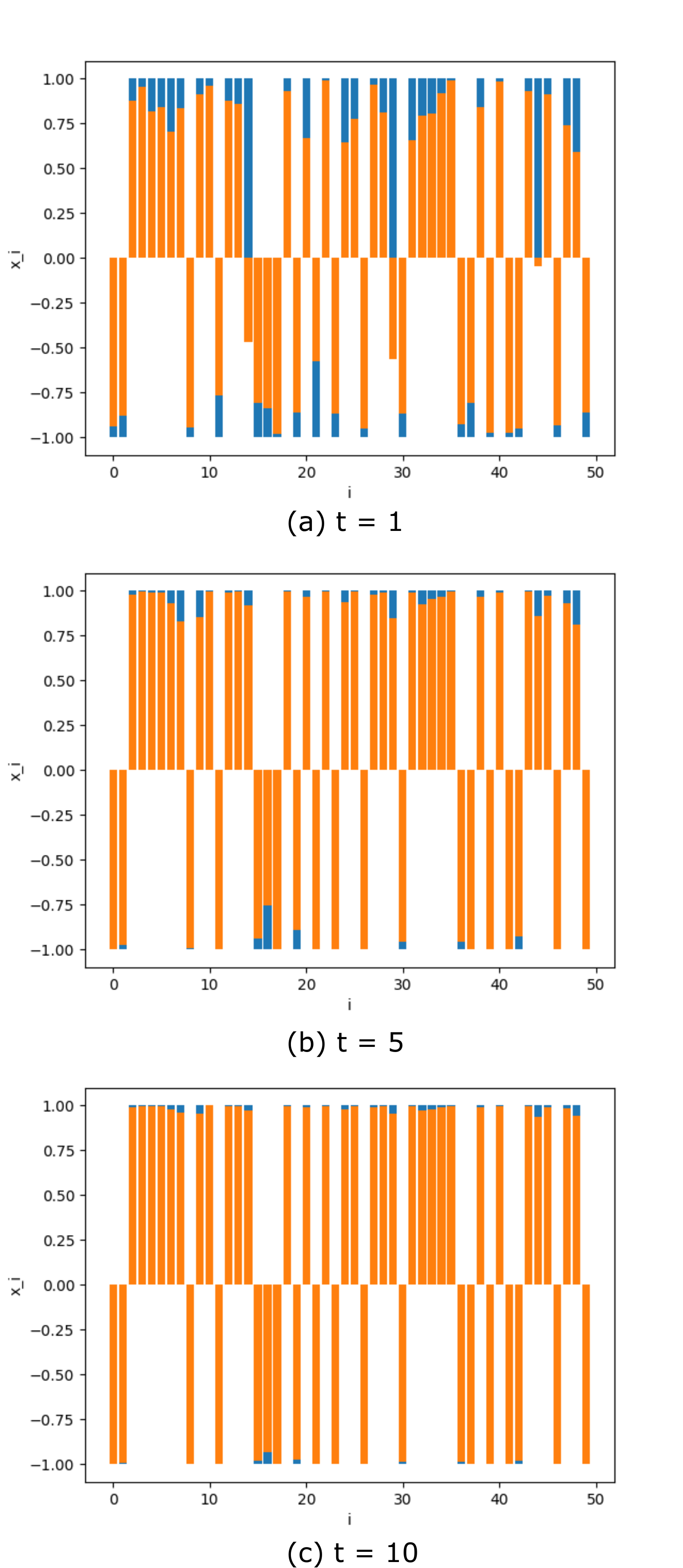}
  {Output $\bm s_{t+1}$ of the TPG-detector (orange symbols) after (a) $t=1$, (b) $t=5$, (c) $t=10$ iterations; $(n,m)=(25,16)$, SNR$=20$ dB, $T=10$.  Blue symbols represent the true signals for comparison.
\label{fig:signal1}  }

We next demonstrate the detection process of the TPG-detector.
Figure~\ref{fig:signal1} shows the output $\bm s_{t+1}$ of the TPG-detector ($T=10$) after $t = 1, 5,$ and $10$ iterations. 
The system size is $(n,m)=(25,16)$ and the SNR is set to $20$ dB.
For comparison, the true signal is also shown in the figure.
We find that the TPG-detector ($t=1$) mistakenly detects a few elements 
of the input signal as shown in Fig.~\ref{fig:signal1} (a).
After the $5$th iteration, as shown in Fig.~\ref{fig:signal1} (b), no bit errors occur while
some values of $\bm{s}_6$ are not $\pm 1$ because of the soft projection in (\ref{eq:hoge_2}).
However, after some additional iterations, all elements of the output are sufficiently close to $\pm{1}$ (see  Fig.~\ref{fig:signal1} (c)) 
indicating that the MSE loss (\ref{eq:sql}) is decreasing.

\subsubsection{BER performances}

\Figure[!t]()[width=0.95\columnwidth]{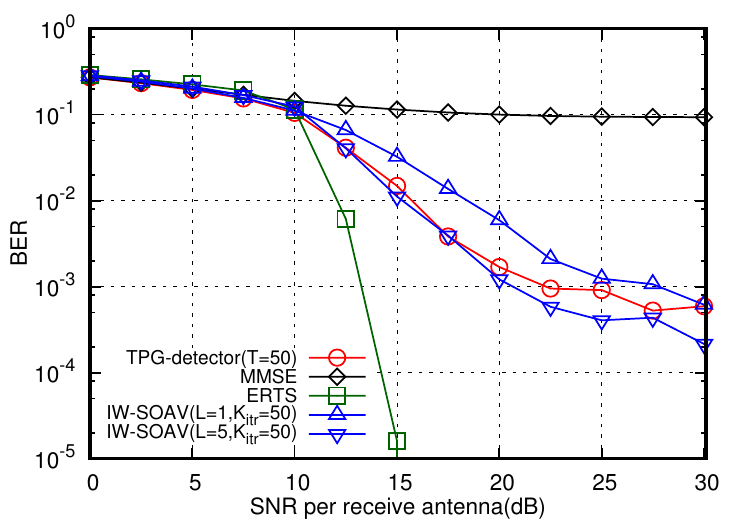}
  {BER performances for {$(n, m) = (50,32)$}.
  \label{fig:BER_N100}}  
  
\Figure[!t]()[width=0.95\columnwidth]{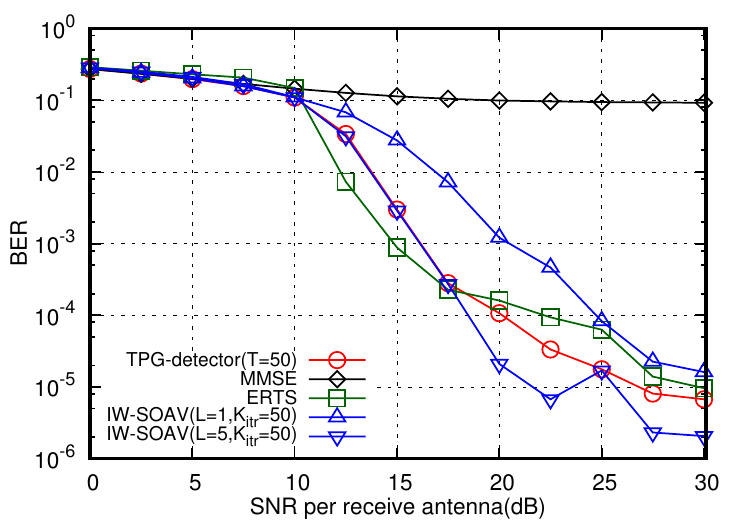}
  {BER performances for {$(n, m) = (100,64)$}.
  \label{fig:BER_N200}}  
  
\Figure[!t]()[width=0.95\columnwidth]{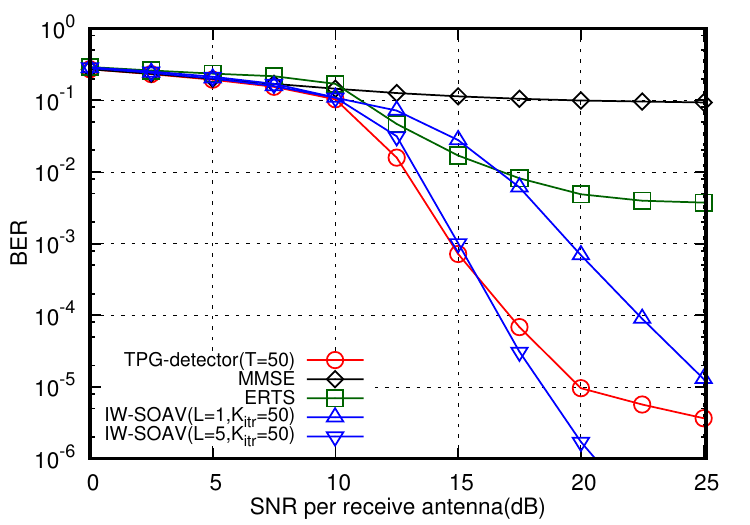}
  {BER performances for $(n, m) = (150, 96)$. \label{fig:BER_N300}}    

We present the BER performance of each detector as a function of SNR for $(n,m)=(50,32)$, $(100,64)$, and $(150,96)$
in Figs.~\ref{fig:BER_N100}-\ref{fig:BER_N300}, respectively.
All the results show that the MMSE detector fails to detect transmitted signals reliably (BER $\simeq 10^{-1}$) 
because the system is underdetermined.

For $(n,m)=(50,32)$ in Fig.~\ref{fig:BER_N100}, {ERTS outperforms the other detection algorithms by a large margin when SNR is larger than $10$ dB.
It should be remarked that ERTS has a much greater time complexity (around several orders of magnitude) than that of the IW-SOAV (see Fig.~7 in \cite{IW-SOAV2}).}
Comparing the TPG-detector with the IW-SOAV, we find that the TPG-detector performs far better than the IW-SOAV ($L=1$) and
shows a BER performance close to the IW-SOAV ($L=5$) when SNR is below 20 dB.
Note that the computational cost for executing {the} TPG-detector with $T=50$ is close 
to that of the IW-SOAV ($L=1$). {The IW-SOAV ($L=5$) requires $K_{\mathrm{itr}}L=250$ iterations, which is 5 times as many as 
the number of iterations required for the TPG-detector with $T = 50$.}
This implies that the TPG-detector can achieve a good detection performance with a relatively low computational cost.

For $(n,m)=(100,64)$ in Fig.~\ref{fig:BER_N200}, 
ERTS detector shows the best BER performance in a middle SNR region 
where SNR is between $10$ and {$18$} dB but the BER curve of ERTS is saturated after $20$ dB.
{The TPG-detector and the IW-SOAV ($L = 1$ and $L=5$) outperform ERTS in a high SNR regime.
In such a regime, the TPG-detector exhibits a BER performance superior to that of the IW-SOAV ($L=1$) for the entire range of SNR,}
i.e., the TPG-detector achieves an approximately $5$ dB gain at $\mathrm{BER}=10^{-4}$ over the IW-SOAV ($L=1$).
More interestingly, the BER performance of the TPG-detector  
is fairly close to that of the IW-SOAV {($L =5$)} when the SNR is below $20$ dB.
When $\mathrm{SNR}=20$ dB, the BER estimate of the TPG-detector  
is $1.0 \times 10^{-4}$ whereas that of the IW-SOAV {($L=5$)}
is $2.1 \times 10^{-5}$.

Figure \ref{fig:BER_N300} shows the BER performance for $(n,m)=(150,96)$.
{In this case, ERTS shows a poor BER performance, and cannot achieve a BER smaller than $10^{-3}$ for any SNR. }
The TPG-detector successfully recovers transmitted signals with lower BER than that of the IW-SOAV ($L=1$).
It again achieves about a $5$ dB gain against the IW-SOAV ($L=1$) at $\mathrm{BER}=10^{-5}$.
In addition, the TPG-detector achieves the lowest BER when $\mathrm{SNR}=12.5$ dB.
Although the IW-SOAV ($L = 5$) shows a considerable performance improvement when $\mathrm{SNR}>15$ dB,
the gap between the curves of the TPG-detector and the IW-SOAV ($L=5$)  {is only } 2 dB at $\mathrm{BER}=10^{-5}$.

\subsubsection{System-size dependency}
{In Fig.~\ref{fig:BER_N}, we show the BER performances of the TPG-detector and IW-SOAV ($L=1$) as a function of
 the number of antennas $n$ with the rate $m/n=0.6$ fixed.
The gap in their BER performances is especially large for SNR$=20$ dB.
We also find that the gain of the TPG-detector increases as $n$ grows despite these algorithms having the same computational costs.
It is confirmed that the TPG-detector outperforms low-complexity algorithms especially in the massive overloaded MIMO channels.}

\Figure[!t]()[width=0.95\columnwidth]{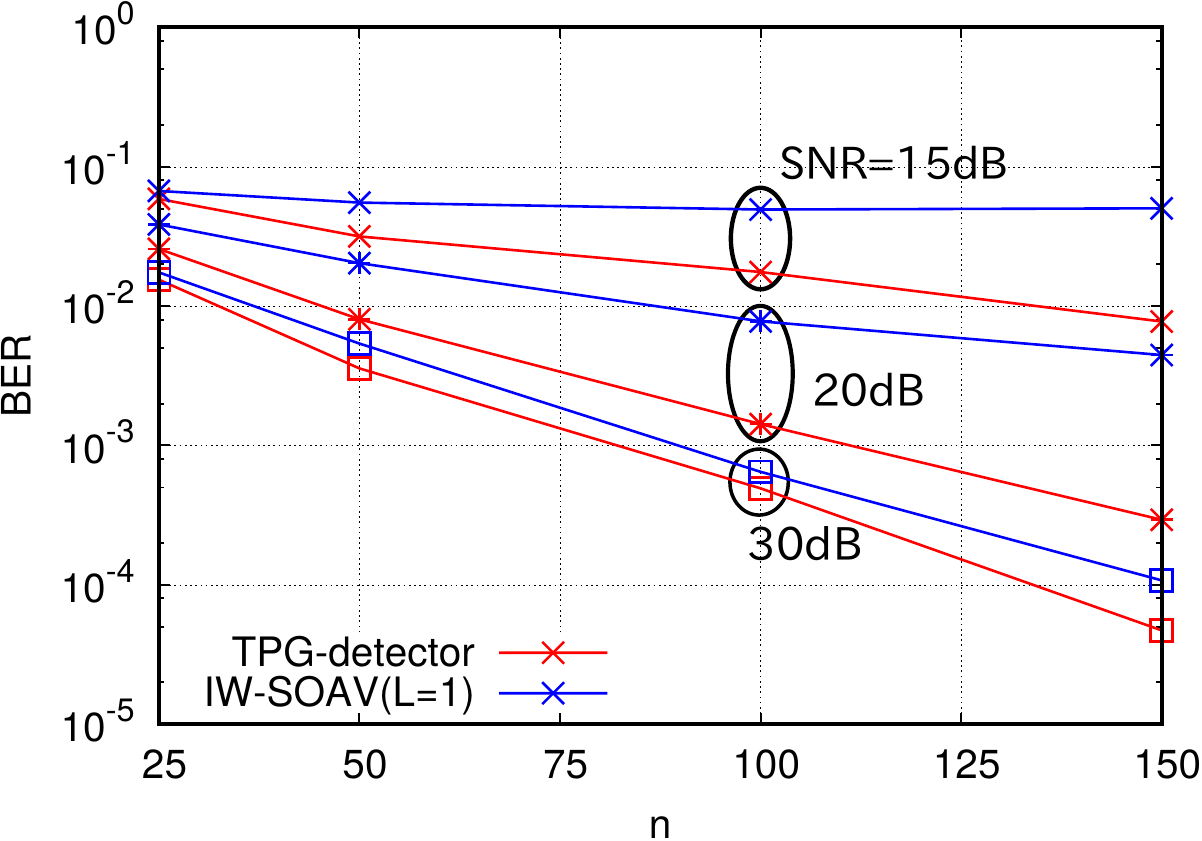}
  {BER performances as functions of the number of antennas $n$ for fixed rate $m/n=0.6$ with SNR$=15,20,30$ dB.
  \label{fig:BER_N}}

\subsubsection{Trained parameters}

Figure~\ref{fig:parameter} displays the learned parameters {$\{\gamma_t\}_{t=1}^T$ and $\{|\theta_t|\}_{t=1}^T$} of the TPG-detector   
after training as a function of the iteration index $t(=1,\dots,T)$.
We find that they exhibit a zigzag shape with a damping amplitude similar to that observed in TISTA~\cite{TISTA}.
The parameter $\gamma_t$, the step size of the linear estimator, is expected to accelerate the convergence of the signal recovery.
Theoretical treatments for providing a reasonable interpretation of 
the characteristic shapes of the learned parameters 
are left as future work.

The trained values of $\alpha$ for different SNR values are shown in Fig.~\ref{fig:parameter_alpha}.
We find that the parameter $\alpha$ depends on the value of SNR.
In particular, the trained value decreases when SNR$\le7.5$ dB.
This tendency is similar to that of another parameter related to $\alpha$ in the IW-SOAV~\cite{IW-SOAV2} using the LMMSE-like matrix.
On the other hand, the trained value is non-monotonic unlike the IW-SOAV; specifically, it increases when SNR$<7.5$ dB.
{The parameter corresponding to $\alpha$ in the IW-SOAV should be chosen in advance by numerical simulations.}
The learning process in the TPG-detector easily tunes the parameter $\alpha$ in addition to other trainable parameters.

\Figure[!t]()[width=0.95\columnwidth]{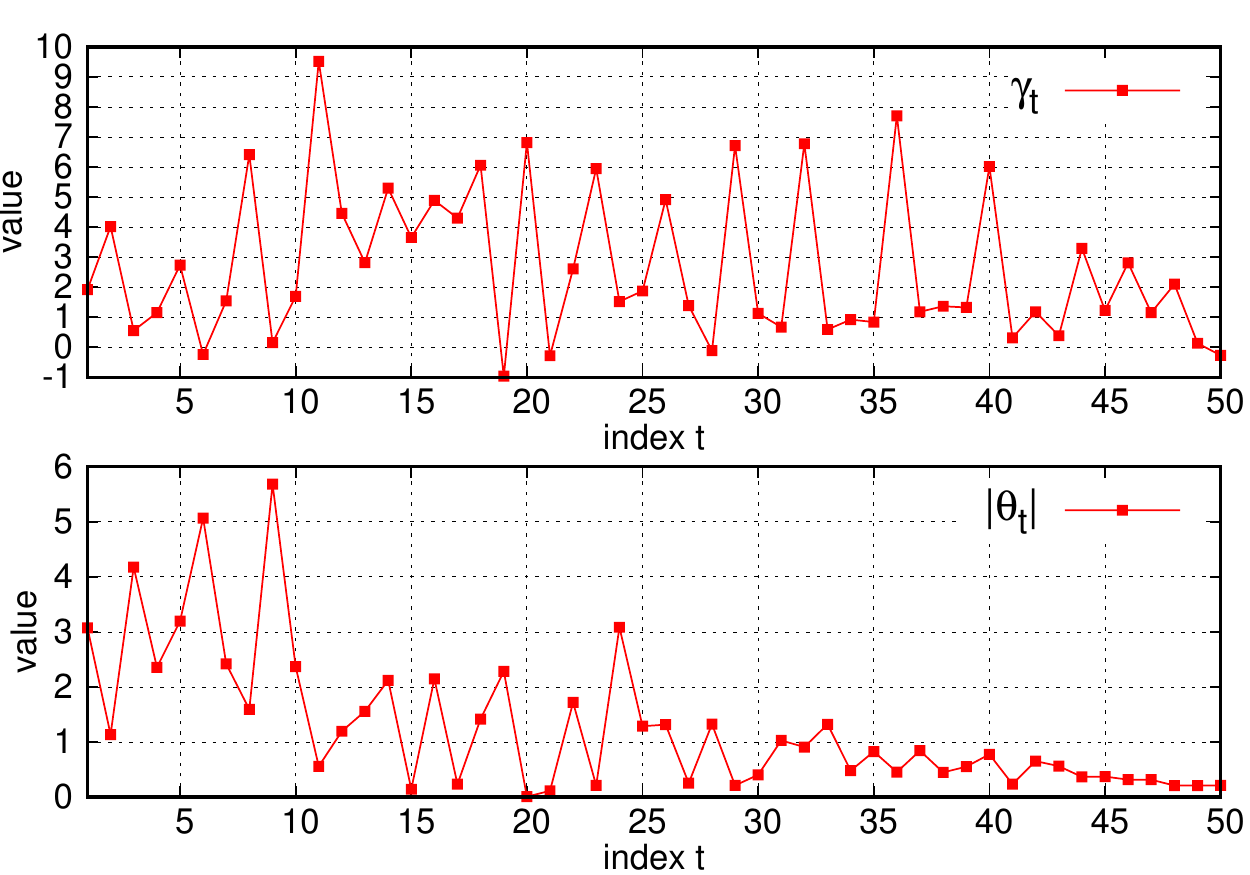}
{Sequences of learned parameters $\gamma_t$ {(top)} and $|\theta_t|$ {(bottom)}; 
  {$(n,m)=(150,96)$}, $\mathrm{SNR}=20$ dB, $1\le t\le T=50$. The trained value of $\alpha$ is $34.68$.  \label{fig:parameter}}

\Figure[!t]()[width=0.95\columnwidth]{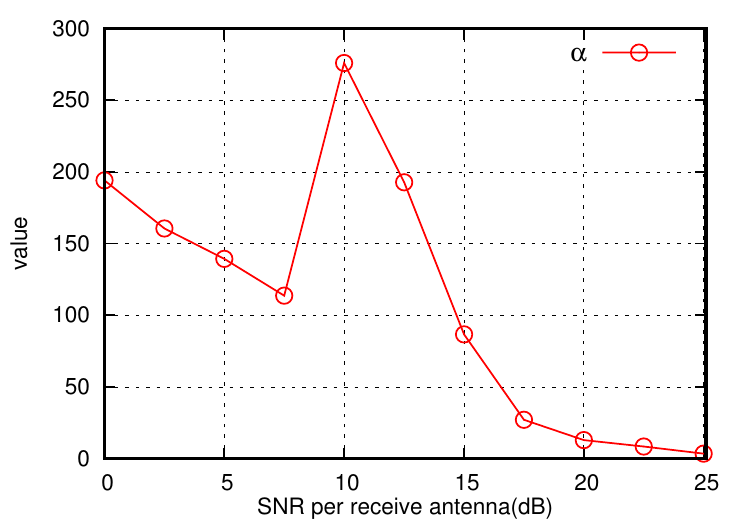}
  {The trained $\alpha$ values for different SNR values; $(n,m)=(150,96)$ and $T=50$.  \label{fig:parameter_alpha}}

\subsubsection{Computation time}

We finally discuss the scalability of the TPG-detector {to show the required computation time for training.}
{The empirical execution time of the training process of the TPG-detector is measured
 by using a PC with GPU NVIDIA GeForce GTX 1080 and Intel Core i7-6700K CPU 4.0 GHz with $8$ cores.
Table \ref{tab:time} presents the execution time of the training processes with different $n$.}
{Even for the case $(n, m) = (150, 96)$, we need only 20 minutes for training the TPG-detector and
 this result indicates that the training process of the TPG-detector is practical for fairly large systems.}

\begin{table}[!t]
  \centering
  \caption{Execution time for training processes of the TPG-detector}
  \begin{tabular}{cccc} \hline
    $(n,m)$ &$(50,32)$&$(100,64)$&$(150,96)$ \\\hline 
    Exec. time (sec) & $638.7$ &$853.2$ & $1203$ \\\hline
  \end{tabular}
  \label{tab:time}
\end{table}

\section{Conclusion}

In this paper, we proposed the TPG-detector, a deep learning-aided iterative decoder for massive overloaded MIMO channels.
It is based on the concept of data-driven {tuning using standard deep-learning techniques}.
{The TPG-detector contains two trainable parameters for each layer: $\gamma_t$ controlling the size of the gradient descent step and
 $\theta_t$ controlling the softness of the soft projection. In addition, the parameter $\alpha$ in the LMMSE-like matrix $\bm{W}$ (\ref{LMMSE_matrix}) is also optimized in the training process.
The total number of trainable parameters in $T$ layers 
is thus $2T + 1$}, which is significantly smaller than used 
in previous studies, such as for the DMD~\cite{DMD1,DMD2}.
This promotes fast and stable training for the TPG-detector.

The computational complexity of the TPG-detector with $T$ iteration steps without the initialization on 
$\bm{W}$ is $O(mnT)$. This is an advantage over the OAMP-net~\cite{OAMP-net} which
needs a matrix inversion for each iteration and has a time complexity of $O(m^3T)$.
This indicates that the TPG-detector is more scalable for massive MIMO systems in terms of the computational cost.

Numerical simulations show that the use of the LMMSE-like matrix successfully 
improves the BER performance of the TPG-detector even when the ratio $m/n$ is relatively large, i.e., 
the channel matrix is ill-conditioned.
It is also revealed
that the proposed TPG-detector outperforms the state-of-the-art IW-SOAV ($L=1$) by a large margin and
achieves a comparable detection performance to the IW-SOAV {($L=5$)}.
The TPG-detector therefore can be seen as a promising iterative detector for massive overloaded MIMO channels 
providing an excellent balance between a low computational cost and a reasonable detection performance.

In this paper, we treat MIMO systems 
with QPSK modulation by separating the real and imaginary part of the signals.
{It is expected that the extension of the TPG-detector to 16QAM or 64QAM modulations is straightforward by using corresponding MMSE functions as a soft projection function
  (see~\cite{TISTA2} for details), which is left for a future work here.
In contrast, this approach will fail for $2^K$-PSK  ($K>2$) modulations}
because it neglects correlations in the modulation format.
Instead, it is crucial to treat complex-valued signals directly as shown in some recent studies~\cite{Jeon, HAYA1}.
As a trainable algorithm based on data-driven tuning, some of the authors recently proposed the 
complex-field TISTA for linear and nonlinear inverse problems in the complex domain~\cite{CTISTA}.
Applying this approach to a massive overloaded MIMO system is a future research task.

\appendices
\section{Brief Review of IW-SOAV}
Here, we give a brief review of {the} IW-SOAV detector.
{The} IW-SOAV is an effective iterative detection algorithm 
for massive overloaded MIMO systems proposed in~\cite{IW-SOAV2, IW-SOAV1}.
It is based on a variant of the Douglas-Rachford algorithm~\cite{RD} which solves the following weighted SOAV (W-SOAV) optimization problem:
\begin{align}
\hat{\bm{s}}\triangleq \mathrm{argmin}_{\bm{z}\in\mathbb{R}^{2n}}
&
\left( \sum_{j=1}^{2n}w_{j}^{+}|{z}_j-{1}|+\sum_{j=1}^{2n}w_{j}^{-}|{z}_j+{1}|\right. \nonumber\\
&\left.
+ \frac{\alpha}{2}\|\bm{y}-\bm{H}\bm{z}\|_2^2\right),
\label{eq_iwsoav0}
\end{align}
where $z_j$ ($j=1,\dots,2n$) is the $j$th element of $\bm{z}$ and $\alpha(>0)$ is a constant.
Here, we assume that each symbol $x_j$ in the transmitted signal $\bm{x}$ 
is an independent random variable which takes a value of $1$ w.p. $w_{j}^{+}$ and $-1$ w.p. $w_j^{-}\triangleq 1-w_j^{+}$.

{The} IW-SOAV repeats the following procedures: (i) estimation of $w_{j}^{+}$ based on the detected signal and
 (ii) detection of the transmitted signal by solving the W-SOAV optimization (\ref{eq_iwsoav0}).
{The} IW-SOAV is thus a double-loop algorithm.

In the outer loop corresponding to procedure (i), the algorithm approximates $\{w_j^{+}\}$ for each transmitted symbol.
The estimation is based on the approximate log likelihood ratio which is given by
\begin{equation}
\hat{\Lambda}_j=\sum_{i=1}^{2m}\frac{2h_{i,j}\{y_i-(\hat{\mu}_i-h_{i,j}\hat{s}'_{j})\}}{\hat{\sigma}_i^2-h_{i,j}^2(1-\hat{s}_j^{\prime 2}) },
\end{equation}
where $h_{i,j}$ is the $(i,j)$th element of matrix $\bm{H}$ and $\bm{\hat{s}}'$ represents a clipped signal of $\bm{\hat{s}}$; specifically i.e.,
$\hat{s}'_j$ ($j=1,\dots,2n$) takes a value of $-1$ if $\hat{s}_j<-1$, and $1$ if $\hat{s}_j>1$, and $s_j$ otherwise.
In addition, we define 
\begin{align}
\hat{\mu}_i &\triangleq \sum_{k=1}^{2n} h_{i,k}\hat{s}'_k,\\
\hat{\sigma}_i^2 &\triangleq \sum_{k=1}^{2n} h_{i,k}^2 (1-\hat{s}_k^{\prime 2}) +\frac{\sigma_w^2}{2}
,\label{eq_iwsoav2}
\end{align}
for $i=1,\dots, 2m$.
Then, the weight $w_j^{+}$ is calculated by
\begin{equation}
w_j^{+} =\frac{e^{\hat{\Lambda}_j}}{1+e^{\hat{\Lambda}_j}}.
\end{equation}

In the inner loop corresponding to procedure (ii), 
the algorithm solves the W-SOAV optimization problem with an iterative process defined by the following
recursive formula:
\begin{align}
\bm{z}_t &= (\bm{I}+\alpha\gamma\bm{H}^{\mathsf T}\bm{H})^{-1}(\bm{r}_t+\alpha\gamma\bm{H}^{\mathsf T}\bm{y})\\
\bm{r}_{t+1} &= \bm{r}_{t} + \theta_t (\phi_{\gamma}(2\bm{z}_t-\bm{r}_t)-\bm{z}_t),\label{eq_iwsoav1}
\end{align} 
where $t(=1,\dots,K_{\mathrm{itr}})$ denotes the iteration step, 
$\theta_t\in [\epsilon,2-\epsilon]$ is a constant, and 
$\phi_{\gamma}:\mathbb{R}^N\rightarrow\mathbb{R}^N$ is a component-wise function 
whose $j$th element $[\phi_{\gamma}(\bm{z})]_j$ is defined by
\begin{align}
[\phi_{\gamma}(\bm{z})]_j \triangleq \begin{cases}
z_j + \gamma & (z_j<-1-\gamma)\\
-1 & (-1-\gamma\le z_j<-1-d_j\gamma)\\
z_j + d_j\gamma & (-1-d_j\gamma\le z_j<1-d_j\gamma)\\
1 & (1-d_j\gamma\le z_j<1+\gamma)\\
z_j - \gamma & (1+\gamma \le z_j)
\end{cases}
,\label{eq_iwsoav2}
\end{align}
with $d_j\triangleq w_j^{+}-w_j^{-}$.
The choice of the parameters $\gamma>0$, $\epsilon\in(0,1)$, and the initial value $\bm{r}_0\in\mathbb{R}^{2n}$ are arbitrary.
In this W-SOAV optimizer, the transmitted symbol is received as $\bm{\hat{x}}=\bm{z}_{K_{\mathrm{itr}}+1}$ 
after $K_{\mathrm{itr}}$ iteration steps.
{The} IW-SOAV starts with $\bm{\hat{s}}=\bm{0}$ and repeats $L$ outer loops with $K_{\mathrm{itr}}$ inner loops.
When all loops are finished, the sign function $\mathrm{sgn}(\cdot)$ is applied to the output $\bm{\hat{x}}$ in an element-wise manner.
The parameter $\alpha$ is fixed appropriately depending on SNR.
In numerical experiments in Section \ref{sec_nr}, we used $\bm{r}_0=\bm{0}$, $\epsilon=0$, $\gamma =1$, and $\theta_t=1.9$ ($t=1,\dots,K_{\mathrm{itr}}$),
and set $\alpha$ as in~\cite{IW-SOAV2}.

The computational cost of each iteration of the IW-SOAV is $O(mn)$.
Although it contains a matrix inversion operation which takes {$O(n^3)$} computational steps, the inversion can be computed in advance.
Since the total number of inner and outer loops is $K_{\mathrm{itr}}L$,
the computational cost of the IW-SOAV without pre-computation is {$O(K_{\mathrm{itr}}Lmn)$}.



\begin{thebibliography}{99}
\bibitem{MUMIMO} E. G. Larsson, O. Edfors, F. Tufvesson, and T. L. Marzetta, ``Massive MIMO for next generation wireless systems,'' IEEE Comm. Magazine, vol. 52, no. 2, pp. 186-195, Feb. 2014.

\bibitem{Yang} S. Yang and L. Hanzo, ``Fifty years of MIMO detection: the road to large-scale MIMOs,'' 
IEEE Comm. Surveys and Tutorials, vol. 17, no. 4, pp. 1941-1988, Fourthquarter 2015.

\bibitem{Shnidman} D. A. Shnidman, ``A generalized Nyquist criterion and an optimum linear receiver for a pulse modulation system,'' The Bell System Technical Journal, vol. 46, no. 9, pp. 2163-2177, Nov. 1967.

\bibitem{SSD} K. K. Wong, A. Paulraj, and R. D. Murch, ``Efficient high-performance decoding for overloaded MIMO antenna systems,''  IEEE Trans. Wireless Commun., vol. 6, no. 5, pp. 1833-1843, May 2007.
\bibitem{ERTS} T. Datta, N. Srinidhi, A. Chockalingam, and B. S. Rajan, ``Low-complexity near-optimal signal detection in underdetermined large-MIMO systems,'' in Proc. National Conf. Comm., Feb. 2012, pp. 1-5.
\bibitem{Fad1} Y. Fadlallah, A. Aïssa-El-Bey, K. Amis, D. Pastor, and R. Pyndiah, ``New decoding strategy for underdetermined MIMO transmission using sparse decomposition,'' in Proc. in Proc. 2013 21st Eur. Signal Proc. Conf., Sep. 2013, pp. 1-5.
\bibitem{IW-SOAV1} R. Hayakawa, K. Hayashi, H. Sasahara, and M. Nagahara, ``Massive overloaded MIMO signal detection via convex optimization with proximal splitting,''
in Proc.  2016 24th Eur. Signal Proc. Conf.,  Aug./Sep. 2016, pp. 1383-1387.
\bibitem{IW-SOAV2} R. Hayakawa and K. Hayashi, ``Convex optimization-based signal detection for massive overloaded MIMO systems,'' in IEEE Trans. Wireless Comm., vol. 16, no. 11, pp. 7080-7091, Nov. 2017.



\bibitem{LISTA} K. Gregor and Y. LeCun,
``Learning fast approximations of sparse coding,''
in {Proc. 27th Int. Conf. Machine Learning}, pp. 399--406, 2010.

\bibitem{ISTA2} I. Daubechies, M. Defrise, and C. De Mol,
``An iterative thresholding algorithm for linear inverse problems with a sparsity constraint,''
Comm. Pure and Appl. Math., vol. 57, no. 11, pp. 1413-1457, Aug. 2004.

\bibitem{TISTA} D. Ito, S. Takabe, and T. Wadayama, ``Trainable ISTA for sparse signal recovery,''
\textit{IEEE Int. Conf. Comm., Workshop on Promises and Challenges of Machine Learning in Communication Networks}, Kansas city, May. 2018. 

\bibitem{TISTA2} D. Ito, S. Takabe, and T. Wadayama, ``Trainable ISTA for sparse signal recovery,''
IEEE Trans. Sig. Process., vol. 67, no. 12, pp. 3113-3125, Jun., 2019.


\bibitem{LP} T. Wadayama and S. Takabe, ``Deep Learning-Aided Trainable Projected Gradient Decoding for LDPC Codes,''
accepted to 2019 IEEE Int. Symp. Info. Theory (ISIT); arXiv:1901.04630.
\bibitem{OAMP-net} H. He, C.-K. Wen, S. Jin, and G. Y. Li,
``A model-driven deep learning network for MIMO detection,'' arXiv:1809.09336, 2018.
\bibitem{CP1} J. Zhang, H. He, C.-K. Wen, S. Jin and G. Y. Li, ``Deep Learning Based on Orthogonal Approximate Message Passing for CP-Free OFDM,'' 2019 IEEE Int. Conf. Acoustics, Speech and Signal Process. (ICASSP), Brighton, United Kingdom, 2019, pp. 8414-8418.

\bibitem{CP2} J. Zhang, C.-K. Wen, S. Jin, and G. Y. Li, ``Artificial Intelligence-aided Receiver for A CP-Free OFDM System: Design, Simulation, and Experimental Test,'' arXiv preprint arXiv:1903.04766 (2019).
\bibitem{SURE} M. Yao, J. Dang, Z. Zhang, and L. Wu, ``SURE-TISTA: A Signal Recovery Network for Compressed Sensing,"
2019 IEEE Int. Conf. Acoustics, Speech and Signal Process. (ICASSP),, Brighton, United Kingdom, 2019, pp. 3832-3836.

\bibitem{DMD1} N. Samuel, T. Diskin, and A. Wiesel, ``Deep MIMO detection,'' 
\textit{2017 IEEE 18th Int. Workshop Signal Processing Advances in Wireless Comm.}, Jul. 2017, pp. 1-5.
\bibitem{DMD2} N. Samuel, T. Diskin, and A. Wiesel, ``Learning to detect,'' IEEE Trans. Signal Process., vol. 67, no. 10, pp. 2554-2564, May. 2019.

\bibitem{Oshea} T. J. O'Shea, T. Erpek, and T. C. Clancy, ``Deep learning based MIMO communications," arXiv preprint arXiv:1707.07980, 2017.
\bibitem{Wang} T. Wang, C. Wen, H. Wang, F. Gao, T. Jiang and S. Jin, ``Deep learning for wireless physical layer: Opportunities and challenges,'' China Comm., vol. 14, no. 11, pp. 92-111, Nov. 2017.
\bibitem{Wen} C. Wen, W. Shih and S. Jin, ``Deep Learning for Massive MIMO CSI Feedback,'' IEEE Wireless Comm. Lett., vol. 7, no. 5, pp. 748-751, Oct. 2018.



\bibitem{TPG} S. Takabe , M. Imanishi, T. Wadayama, and K. Hayashi, ``Deep Learning-Aided Projected Gradient Detector for Massive Overloaded MIMO Channels,''  accepted to IEEE International Conference on Communications (ICC2019), 2019; https://arxiv.org/abs/1806.10827.

\bibitem{PIC} D. Divsalar, M. K. Simon, and D. Raphaeli, 
``Improved parallel interference cancellation for CDMA'', IEEE Trans. Commun., vol. 46, no. 2, pp. 258-268, Feb. 1998.

\bibitem{Adam} D. P. Kingma and  J. L. Ba, ``Adam: A method for stochastic optimization,'' arXiv:1412.6980, 2014.

\bibitem{OAMP} J. Ma and L. Ping,
``Orthogonal AMP,''
IEEE Access, vol. 5, pp. 2020-2033, Jan. 2017.

\bibitem{PyTorch}
PyTorch, \url{https://pytorch.org}.



\bibitem{Jeon} C. Jeon, R. Ghods, A. Maleki, and C. Studer, ``Optimality of large MIMO detection via approximate message passing," 2015 IEEE International Symposium on Information Theory (ISIT), Hong Kong, 2015, pp. 1227-1231.

\bibitem{HAYA1} R. Hayakawa and K. Hayashi, ``Reconstruction of complex discrete-valued vector via convex optimization with sparse regularizers," IEEE Access, vol. 6, pp. 66499-66512, Oct. 2018.


\bibitem{CTISTA} S. Takabe and T. Wadayama, ``Complex field-trainable ISTA for linear and nonlinear inverse problems,'' arXiv:1904.07409, 2019.

\bibitem{RD} P. Combettes and J.-C. Pesquet, ``Proximal splitting methods in signal processing,'' 
in Fixed-Point Algorithms for Inverse Problems in Science and Engineering 
(Springer Optimization and Its Applications), vol. 49. New York, NY, USA: Springer, 2011, pp. 185-212.

\end{thebibliography}

\EOD
\end{document}